\documentclass[11pt,a4paper]{emulateapj}
\usepackage{natbib}
\usepackage{amssymb, amsmath, amsbsy, epsfig, epsf, slashbox}
\usepackage{color}

\slugcomment{Accepted for publication in {\it The Astrophysical Journal}}
\begin{document}

\title{A Glimpse at Quasar Host Galaxy Far-UV Emission, using DLAs as Natural Coronagraphs}
\author{Zheng Cai\altaffilmark{1,2}, Xiaohui Fan\altaffilmark{1}, Pasquier Noterdaeme\altaffilmark{3}, Ran Wang\altaffilmark{1}, Ian McGreer\altaffilmark{1}, Bill Carithers \altaffilmark{4}, Fuyan Bian\altaffilmark{5,13}, Jordi Miralda Escude\altaffilmark{6}, Hayley Finley\altaffilmark{3}, Isabelle P$\hat{a}$ris\altaffilmark{7}, Donald P. Schneider\altaffilmark{8, 9}, Nadia L. Zakamska\altaffilmark{12}, Jian Ge\altaffilmark{10}, Patrick Petitjean\altaffilmark{3}, Anze Slosar\altaffilmark{11}}
\affil{Steward Observatory, University of Arizona, Tucson, AZ 85721}
\affil{Physics Department, University of Arizona, Tucson, AZ 85721}
\affil{Institut d'Astrophysique de Paris, CNRS-UPMC, UMR7095, 98bis bd Arago, 75014 Paris, France}
\affil{Lawrence Berkeley National Laboratory, 1 Cyclotron Road, Berkeley, CA, USA}
\affil{Research School of Astronomy and Astrophysics, Australian National University, Canberra, Weston Creek, ACT, 2611}
\affil{Institut de Ci\`encies del Cosmos, Universitat de Barcelona, Spain}
\affil{Departamento de Astronomia, Universidad de Chile, Casilla 36-D}
\affil{Department of Astronomy and Astrophysics, The Pennsylvania State University, University Park, PA 16802}
\affil{Institute for Gravitation and the Cosmos, The Pennsylvania State University, University Park, PA 16802}
\affil{Department of Astronomy, University of Florida, USA}
\affil{Brookhaven National Laboratory, Upton, NY 11973, USA}
\affil{Department of Physics \& Astronomy, Johns Hopkins University, 3400 N. Charles St., Baltimore, MD 21218, USA}
\affil{Stromlo Fellow}


\altaffiltext{1} {Email: caiz at email.arizona.edu}
%
%

\begin{abstract}
In merger-driven models of massive galaxy evolution, the luminous quasar phase is expected to be 
accompanied by 
vigorous star formation in quasar host galaxies. In this paper, we use high column density Damped Lyman Alpha 
(DLA) systems along quasar sight lines as natural coronagraphs to directly study the far-UV (FUV) radiation from the host galaxies of 
luminous background quasars. We have stacked the spectra of $\sim$2,000 DLA systems ($N_{\rm{HI}}>10^{20.6}$ cm$^{-2}$) with a median 
absorption redshift $\left\langle z \right\rangle=2.6$ selected from quasars observed in the SDSS-III Baryon Oscillation Spectroscopic Survey. We detect 
residual flux in the dark troughs of the composite DLA spectra. The level of this residual flux significantly exceeds systematic 
errors in the SDSS fiber sky subtraction; furthermore, the residual flux is strongly correlated with the continuum luminosity of 
the background quasar, while uncorrelated with DLA column density or metallicity. We conclude that the flux could be associated with the 
average FUV radiation from the background quasar host galaxies (with medium redshift $\left\langle z\right\rangle=3.1$) that is not blocked by the intervening DLA. Assuming all of the detected flux originates from quasar hosts, for the highest quasar luminosity bin 
($\left\langle L\right\rangle=$2.5$\times 10^{13} L_\odot$), the host galaxy has a FUV intensity of 
$1.5 \pm 0.2 \times 10^{40}$ erg s$^{-1}$ \AA$^{-1}$; this corresponds to an unobscured UV star formation rate of 
9 M$_\odot$ yr$^{-1}$. 

\end{abstract}

\section{Introduction}



It has been shown that, in the local universe, most luminous and massive spheroidal galaixes have central super-massive black holes (SMBHs) \cite[e.g.][]{mclure99}, and that fundamental relationships exist between the black hole mass and bulge stellar mass \cite[e.g.][]{magorrian98, ferrarese00, gebhardt00, marconi03, haring04}. 
Likelywise, many high-redshift quasars are associated with massive host galaxies \cite[e.g.][]{aretxaga98, carilli01}. The study of high redshift quasar host galaxies opens up an important avenue to study the assembly and evolution of massive galaxies, in particular the relationship to the nuclear black holes \cite[e.g.][]{schramm08}. 

In the merger-driven evolutionary model, 
quasars are triggered by interactions or mergers of gas-rich galaxies \cite[e.g.][]{sanders88, hopkins06}. The interaction of galaxies produces inflows of gas, which simultaneously fuels both intense star formation in the host galaxies and black hole growth \citep{hopkins06}. The optically luminous quasar phase follows a dusty ultra-luminous infrared galaxy (ULIRG) phase and a dust-ejection phase 
 (e.g., Hopkins et al. 2006). The study of SFR and dust obscuration in quasar hosts would provide strong tests    of this merger-driven evolutionary model, and cast light on the relationship between quasar hosts and ULIRGs. However, progress in this field has been hindered by the difficulty of isolating quasar host galaxies, since such observations are invariably hindered by the immense brightness of the quasar nuclei in the rest-frame UV/optical regime. 

Observations at millimeter (mm) and sub-mm wavelengths, which correspond to the rest-frame far-IR (FIR), are crucial for probing high-redshift quasar hosts, because the FIR signal traces the large molecular gas reservoirs which fuel star formation \cite[e.g.][]{carilli06}, and the contribution from the star formation can be disentangled from the AGN activity in the FIR regime \cite[e.g.][]{yun00}. Recent observations with the Atacama Large Millimiter/submillimeter Array (ALMA) suggest that the $z\sim 6$ quasar hosts have dynamical masses  $\rm{M}_{\rm{dyn}}=10^{10}- 10^{11}$ M$_\odot$, one order of magnitude higher than that of typical local galaxies \citep{wang13}; and that the star formation rate (SFR) of the quasar hosts ranges from a few hundred to one thousand M$_{\odot}$ yr$^{-1}$ \cite[e.g.][]{wang11}. 


At shorter wavelengths, recent surveys of quasar hosts have concentrated on the near-infrared, at 
rest-frame wavelengths greater than the Balmer break, and where the fraction of the observed light due to the host galaxy is larger than that in the rest-frame UV \cite[e.g.][]{guyon06}. Hubble Space Telescope (HST) and ground-based observations with excellent seeing conditions ($<0.4\rq{}\rq{}$) or near-IR adaptive optics (AO) have been used to successfully detect the rest-frame optical light of the quasar host galaxies \citep{kukula01, ridgway01, peng06, targett13, young13}. However, at $z>1$,  obtaining reliable measurements of the host galaxy luminosities remains extremely difficult, because of the severe effect of surface brightness dimming at high redshift and the technical difficulties associated with the construction of a precise PSF \citep[e.g.][]{hutchings02, targett13}. 

There has been much interest in directly tracing the formation of massive young stars in high redshift galaxies in the UV \cite[e.g.][]{shapley05, bian12}. However, it is even more difficult to resolve and detect the high redshift quasar hosts at the rest-frame UV regime with current generation telescopes, especially in the far-UV (FUV) regime ($\lambda \sim 1000$\AA). The contrast between the bright nuclear point source and the host galaxy increases dramatically beyond $z>1$, and the FUV light from host galaxies is inevitably swamped by the UV-bright nuclei. Reliable measurements of the FUV emission of the host galaxies are extremely difficult using the traditional PSF subtraction method. Recent deep HST observations did not detect the host galaxy of one $z=6$ quasar in the rest-frame near-UV regime, even with delicate observations designed to minimize PSF variations and provide careful PSF subtraction \citep{mechtley12}. New techniques, such as coronagraphs, must be developed to block out the UV light from central AGN, and in doing so to reveal the UV emission from host galaxies \cite[e.g.][]{martel03, finley13}. 

Damped Ly$\alpha$ absorption (DLAs) systems at the redshift of the quasar, hereafter 
called associated DLAs, have recently been developed as a technique for probing quasar 
hosts in the UV regime \citep{hennawi09, zafar11}. The idea is 
that some associated DLAs completely absorb the strong Ly$\alpha$ emission from the 
central AGN, but do not fully obscure the larger-scale extended Ly$\alpha$ emission or 
Lyman continuum from the host galaxy. Therefore, associated DLAs can act as natural 
coronagraphs and enable the study of the Ly$\alpha$ emission from the quasar hosts 
(Finley et al. 2013). Utilizing long-slit spectroscopy, Zafar et al. (2011) reported the detection 
of the Lyman continuum of the host galaxy in the dark trough 
of the DLA in front of quasar Q0151+048A at $z=1.9$.

In this paper, we demonstrate a novel techique for probing the FUV emission from quasar host galaxies. By stacking a 
large number of intervening DLAs in the foreground of quasars from the SDSS-III Baryon Oscillation Spectroscopic Survey (BOSS), 
we are able to detect the FUV emission of the quasar hosts within the stacked DLA dark trough. The HI clouds in 
galactic or circumgalactic environments are likely to be clumpy and the HI volume filling factor is expected 
to be significantly smaller than unity \cite[e.g.][]{braun92, borthakur10}. Therefore  
the clumpy HI gas may only block the quasar continuum, whereas much of the host remains unobscured. 
In \S 2, we discuss the observations and our DLA sample selection. The results of the composite spectra are presented in \S 3. 
We discuss the implication of our results on the origins of the flux in the composite DLA trough in \S 4. 
Throughout this paper, we adopt a cosmology based upon the fifth year Wilkinson Microwave Anisotropy Probe (WMAP) data \citep{komatsu09}: $\Omega_{\Lambda}=0.72$, $\Omega_m= 0.28$, $\Omega_b=0.046$, and $\rm{H}_0=70\ \mbox{km}\ \rm{s}^{-1}\ \mbox{Mpc}^{-1}$. 


\section{Method and Sample}

\subsection{Using intervening DLAs as Coronagraphs}

DLAs are characterized by a wide flat trough with zero transmitted flux (i.e., dark trough or dark core), resulting from the natural broadening of the absorption at high HI column densities. In this study, we define the DLA dark trough as a region where the flux is negligibly small; specifically, regions where the ratio of the expected flux to the continuum flux is $< 1\times 10^{-4}$. Intervening DLAs can act as natural coronagraphs, because the quasar continuum, emitted from the centrally bright nucleus, is completely absorbed within the DLA dark trough. However, it is quite likely that regions of the 
extended emission from the quasar hosts is unobstructed by the intervening clumpy HI absorbers, and the extended stellar light thus leaks through and can be detected within the dark trough of the foreground DLAs.

\subsection{The Clean DLA (CDLA) sample}

Our sample is selected from the Baryon Oscillation Spectroscopic Survey (BOSS) \citep{dawson13, ahn13}. 
BOSS is a spectroscopic survey that will measure redshifts of 1.5 million luminous red galaxies and 
Ly$\alpha$ absorption towards 160,000 high redshift quasars (Eisenstein et al. 2009). 
1,000 optical fibers with 2 arcsec diameters are plugged into the 1,000 holes in an aluminum 
plate to receive light from 160-200 quasars, 560-630 galaxies, ~100 ancillary science targets and standard stars (e.g., Dawson et al. 2013). 
Also, each plate contains at least 80 sky ﬁbers which placed at blank area on the sky to 
model the background for all the science fibers. 
The distribution of the sky fibers is constrained to cover the entire focal plane. 
This designs allow the sampling of the varying sky background over the 
focal plane the spectrograph (e.g., Dawson et al. 2013). 
The BOSS spectra have the moderate resolution of $R\sim 2000$ from 3700 \AA- 10,400 \AA. 
With the typical exposure time of 1-hour for each plate, the BOSS spectra have a 
modest median S/N of $\sim2-3$ per resolution element for quasars at rest-frame 
wavelength $\lambda = 1041- 1185$ \AA (Lee et al. 2012).

The Data Release 10 Quasar (DR10Q) catalog \citep{paris13} includes 166,798 quasars detected over 6,000 deg$^2$, of which 117,774 are at $z>2.15$. A combination of target selection metods \citep{ross12a, bovy11} achieved this high surface density of high-redshift quasars. Using a fully automated procedure based on profile recognition using correlation anaysis, \citet{noterdaeme12} constructed a catalog of DLAs and sub-DLAs with 
$N_{\rm{HI}} > 10^{20.0}$ cm$^{-2}$ detected from an automatic search along DR9Q lines-of-sight. Using a similar technique, \citet{noterdaeme14} generated the DLA catalog for Data Release 10 (DR10). 

The DLA detection technique \citep{noterdaeme12, noterdaeme09} pays special attention to the low-ionization metal lines in order to remove 
possible blends or mis-identifications. The redshift measurement is mainly based on the profile fit the DLA. Some of the DLA redshift measurements have been refined based on the low-ionisation metal lines. The equivalent width of the metal lines is obtained automatically by locally normalising the continuum around each line of interest and then modeling the absorption lines with a Voigt profile. For non-detections, upper limits are obtained from the noise around the expected line positions. 

The final BOSS DR10 DLA catalog includes 11,030 DLAs with $N_{\rm{HI}} > 10^{20.3}$ cm$^{-2}$, continuum-to-noise ratio (CNR) $> 2$, and absorption redshift $z_{\rm{abs}}>2.15$. We avoided proximate DLAs at velocities less than 5,000 km s$^{-1}$ from the quasar as these have a high 
probability of being physically associated with the local quasar environment. Strong proximate DLAs from BOSS are studied in Finley et al. (2013).

In our investigation, the DLA purity is crucial. Guided by this consideration, we include only DLAs satisfying the following stringent criteria: (a) spectra with median CNR $> 4.0$; (b) $>3\sigma$ low-ionization metal line detections; (c) column densities $N_{\rm{HI}}>10^{20.6} $ cm$^{-2}$; which requires the width of the DLA trough to be greater than four times the spectral resolution; (d) Spearman\rq{}s correlation coefficient of the Voigt profile fitting is greater than 0.6, which ensures an accurate estimate of the DLA column density (see details in Noterdaeme et al. 2009). Criterion (b) ensures that it is a DLA that is responsible for absorption rather than an unresolved blend of multiple narrow absorption features, while 
criteria (a), (c) and (d) enable an accurate determination of the column density and redshift. 
$\sim$2,300 out of 11,030 DLAs in the catalog satisfy these four criteria. 
Further, we visually inspect the 2,300 systems to confirm that they are DLAs. We eliminate $\sim 40$  systems close to quasar OVI+Ly$\beta$ region. Also, the visual inspection enables us to find a small number of DLAs with inaccurate redshift measurements, and we paid careful attention to refine the redshifts for some of the DLAs according to Voigt profile fitting and metal lines if necessary. We also drop a number of 
 DLAs with blending of Lyman limit systems or strong Ly$\alpha$ forest absorption in both DLA wings, which makes the measurements of the DLA column density and redshift difficult. A sample of 2,138 DLAs pass the visual inspection. Further, we constrain the redshift of DLAs in the moderate redshift range of $2.2<z<3.2$. 
 A sample of 1,940 DLAs within the redshift range are finally selected and defined. In the remainder of the paper, we refer to these 1,940 DLAs as the {\it clean DLA sample} ({\it CDLA} sample). 
In Fig.\enskip1, the right panel shows the redshift distribution of these 1,940 DLAs with different quasar luminosity bins. The orange histogram represent the highest luminosity bins, while the red histogram represents the lowest luminosity bin. The left panel shows the redshift distribution for the corresponding background quasars. The redshift distribution of DLAs in different luminosity bins are similiar within the redshift range of $2.2<z<3.2$.

\figurenum{1}
\begin{figure}[tbp]
\epsscale{1}
\label{fig:images}
\plotone{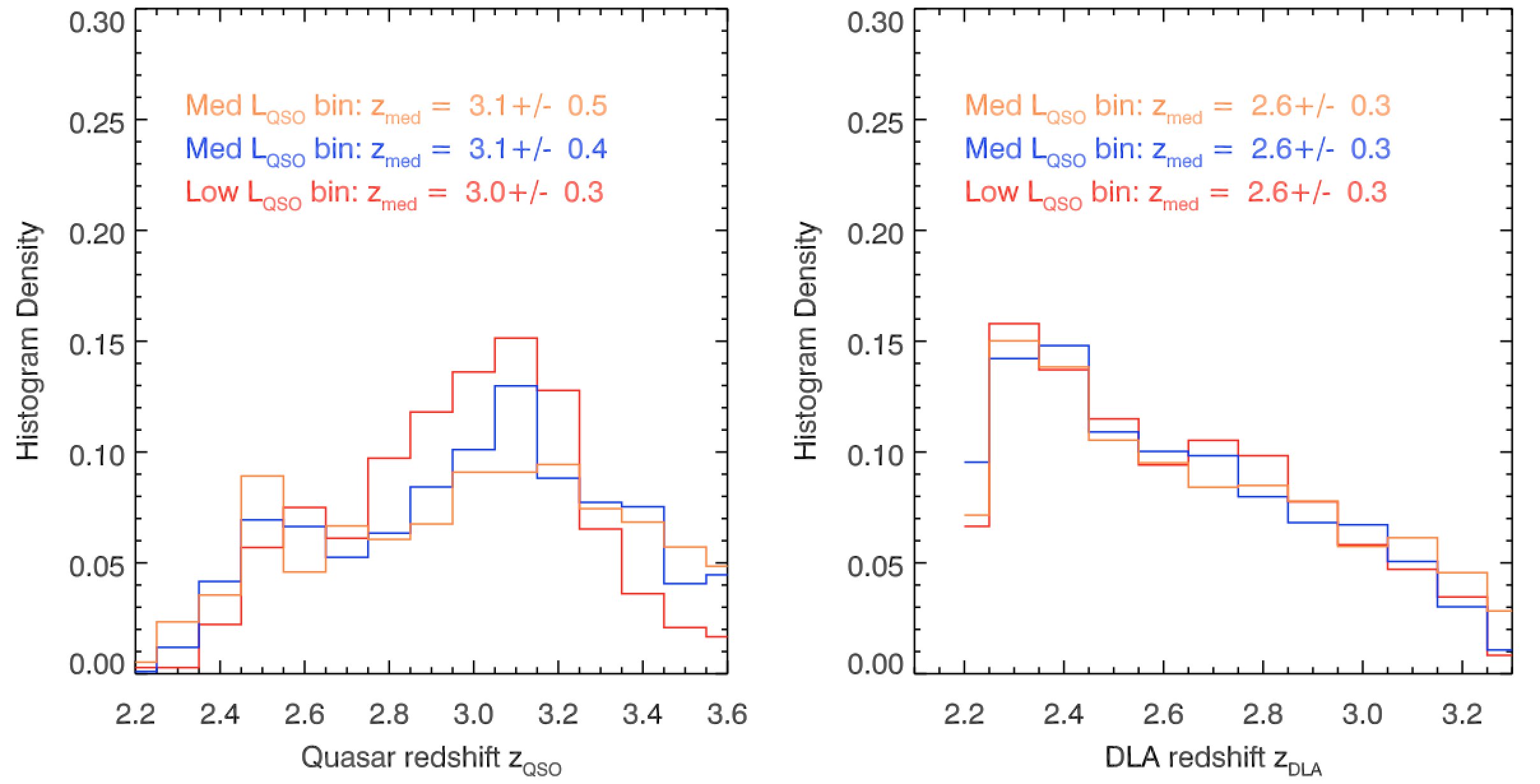}
\caption{The right panel shows the redshift distribution of SDSS DR10 DLAs with different quasar luminosity bins. 
The orange histogram represents the highest quasar luminosity bin, while the red histogram represents the lowest quasar luminosity bin. 
The left panel presents the redshift distribution of the background quasars. The DLA redshift range we consider is $z=2.2-3.2$. The redshift distribution of the DLAs and corresponding background quasars in different quasar luminosity bins are similiar in the 
DLA redshift range of $z=2.2-3.2$. }
\end{figure}

DLAs with $N_{\rm{HI}}> 10^{20.6}$ cm$^{-2}$ have a dark trough with a rest-frame velocity width extending at least $\pm 340$ km s$^{-1}$ 
from the center, much larger than the DLA redshift uncertainties.
In addition to refining the DLA redshifts from Voigt profile fitting on Ly$\alpha$ part, we use multiple low-ionization metal lines to fit the redshifts ($z_{\rm{ion}}$) of $ \sim 500$ DLAs. We find that the largest difference between the redshift fitted by low-ionization metal lines and redshift fitted by Ly$\alpha$ Voigt profile ($c\times \Delta z/(1+z))$ is $\sim 150$ km s$^{-1}$.
We also checked the DLA trough of the composite DLA 
spectra (see \S3.1 and \S3.3, and also the inset in Fig.\enskip3), and the central $\pm 250$ km s$^{-1}$ is 
consistent with a flat absorption trough within 1-$\sigma$.  
All of this evidences support that the DLA redshift uncertainties are much smaller than the width of DLA dark trough; and we conservatively define the dark trough in the composite spectra as the region within $\pm 150$ km s$^{-1}$ about the DLA center.

\section{Residual Flux in the Dark Trough}

\subsection{Initial stacking}

The SDSS spectra are shifted to the rest-frame of the DLA while conserving the flux, i.e., $f_{\rm{i}}(\lambda_{\rm{res}})= (1+z_{\rm{DLA}})   \times f_{\rm{i,sdss}}(\lambda_{\rm{obs}})$, where $\lambda_{\rm{res}}$ is the DLA rest-frame wavelength; $\lambda_{\rm{obs}}$ is the wavelength in the SDSS spectra ($\lambda_{\rm{obs}}= \lambda_{\rm{res}}\times (1+z_{\rm{DLA}}$)); 
 $f_{ \rm{i}}(\lambda)$ is the flux density of an individual spectrum redshifted to the rest frame; and $f_{\rm{i, sdss}} (\lambda)$ is the flux density of an individual SDSS spectrum in the observed frame. In the dark trough, the flux density 
in each pixel (wavelength) of the composite spectra is calculated by two methods: (1) the 3$\sigma$-clipped mean of all the spectra at the same wavelength; (2) median of all the spectra at the same wavelength. Outside the dark trough with $\lambda < 1213$\AA\ or $\lambda > 1218$\AA, we take the median value to stack the spectra. We calculate the error at each pixel (wavelength) in the stacked spectrum by propagating the errors of the pixels at the same position in every individual spectrum $\left(\sigma_{\rm{stacked, i}}= \frac{1}{n}\times \sqrt{\sum \limits_{i=0}^n \sigma^2_{i}} \right)$. Note that 3$\sigma$-clipping mainly removes the large outliers due to noise, but not clipping away the Ly$\alpha$ emission from DLA galaxy. On the one hand, even assuming most of DLA host galaxies are as bright as $L^*$ galaxy at $z=2-3$ (e.g., Shapley et al. 2003), the Ly$\alpha$ emission from DLA host is still submerged to the noise of the individual SDSS-III spectrum. The detailed discussion is included in \S 4.1.  On the other hand, from visual inspection of the 
entire {\it CDLA} sample, we do not find DLAs with strong Ly$\alpha$ emission in the dark trough. 

In the stacking process, we do not introduce any additional scaling.  
It is true that different quasars are different in luminosity, quasar
host flux, and DLA host emission. However, in the DLA dark trough, quasar continuum is completely 
blocked in the DLA dark trough. Also, for SDSS-III spectra, the average flux densities of the DLA hosts and quasar hosts 
are generally about or more than one order of magnitude lower than the noise level 
(see more details in \S4.1 and \S4.2). For SDSS-III spectrum, the noise is mainly due 
to the sky background and the CCD read noise (e.g., Dawson et al. 2013), and in the dark trough of DLAs at $z=2.6\pm 0.3$, 
 different systems generally have similar noise level. 
Therefore, following previous similar studies, we just simply stacking the spectra without
any additional scaling (e.g., Rahmani et al. 2010; Noterdaeme et al. 2014).

After the initial stacking, we find that the dark trough of the stacked DLA shows a positive offset.  
We calculate the mean flux density by averaging the flux density in the dark trough region. Following the discussions in \S 2.2, we conservatively define the stacked dark trough region within $\pm$150 km s$^{-1}$ about the center. For the stacking of 3$\sigma$-clipped mean, the average dark trough flux density is $\bar{F}_{\rm{dark}}=6.5 \pm 0.6 \times 10^{-19} $ ergs cm$^{-2}$ A$^{-1}$ s$^{-1}$; and for the median stacking, the average dark trough flux density is $\bar{F}_{\rm{dark}}=7.0 \pm 0.6 \times 10^{-19} $ ergs cm$^{-2}$ A$^{-1}$ s$^{-1}$.  A general flux residual in the stacked DLA dark trough has been documented in previous work with much smaller sample size \citep{rahmani10, paris10}. The large DLA database of SDSS-III/BOSS DR10 enables us, for the first time, to carefully study and test the origin of the positive residual flux in the dark core. This residual flux can arise from three sources: (a) sky subtraction residual (systematic sky subtraction errors); (b) Ly$\alpha$ emission around $\lambda \sim$1216 \AA\ from the DLA galaxies; (c) FUV continua emission at $(1+z_{\rm{DLA}})/(1+z_{\rm{QSO}}) \times 1216$\AA\ $\sim$ 1100 \AA\ from the quasar host galaxies which is not blocked by the foreground DLAs. 

\subsection{The determination of the sky subtraction residual}

 In order to calculate the sky subtraction residual, we select a group of 150 DLAs at $z>3.8$ with a median redshift $<z>= 4.0$, and examine the flux in the DLA Lyman limit region ranging from rest-frame $\lambda=800$ \AA\ to $900$ \AA, where the optical depth $\tau(\lambda)$ is expected to be over $1000$. For the sample of $z>3.8$ DLAs, the Lyman limit region at rest-frame $\lambda=820 $\AA\ to $900$\AA\ corresponds to an observed wavelength from $ 4100$\AA\ to $4500$\AA, which covers the average wavelength of DLA trough in our {\it CDLA} sample. 
 The expected flux of quasar continua should be negligible small in the DLA Lyman limit region because of the large optical depth. The flux in this region can only be contributed by: (a) sky-subtraction residuals; (b) unobscured escaping ionizing photons with $\lambda \lesssim 900$\AA\ from DLA galaxies and (c) photons from quasar hosts at even shorter wavelength. Therefore, the average flux in DLA Lyman limit regions can be regarded as an upper limit for the sky-subtraction residual. A number of studies have found that the escape fraction is small, $\lesssim$ 5\%- 10\% relative to photons escaping at 1500 \AA\ \citep{malkan03, siana07, bridge10}. This small escape fraction suggests that the ionizing photons could only have a minor or even negligible contribution to the average flux in the DLA Lyman limit region. 

The 150 DLAs at $z>3.8$ we have selected from the DLA catalog all have CII and/or SiII detections and passed the visual inspection to ensure the DLA 
nature. We use the median to stack the spectra at rest-frame $\lambda = 820$ - $900$\AA\ in the stacked spectra. Then, we average the flux from $\lambda = 820$\AA - $900$\AA. The average flux density is equal to $3.502 \pm 0.008 \times 10^{-19} $ ergs cm$^{-2}$ A$^{-1}$ s$^{-1}$, and this value can be regarded as an upper limit of the sky-subtraction (see red dot in Fig.\enskip2).  
We double checked our results using the average sky subtraction residual as a function of wavelength which is determined by the SDSS pipeline group (\citealp{bolton12} (Schlegel et al. 2014 in prep., private communication). The sky-subtraction residual is $\sim 3\times 10^{-19}$ ergs cm$^{-2}$ A$^{-1}$ s$^{-1}$ at $\sim 4300$\AA, consistent with the upper-limit on the sky-subtraction residual we have derived. In Fig.\enskip2, the inset shows a zoom-in of the DLA absorption for the composite spectrum of the group of $z\ge 3.8$ DLAs. Also, in Fig.\enskip2, we present the median flux in the composite DLA dark trough (blue dot with errorbar) of the {\it CDLA} sample is $6.5\pm0.6 \times 10^{-19}$ ergs cm$^{-2}$ \AA$^{-1}$ s$^{-1}$, which is significantly larger than the SDSS sky-subtraction residual (red dot in Fig.\enskip2). The observed wavelength of blue dot in Fig.\enskip2 is determined by the center of DLA trough at the median redshift of CDLA sample.

\figurenum{2}
\begin{figure}[tbp]
\epsscale{1}
\label{fig:testschemes}
\plotone{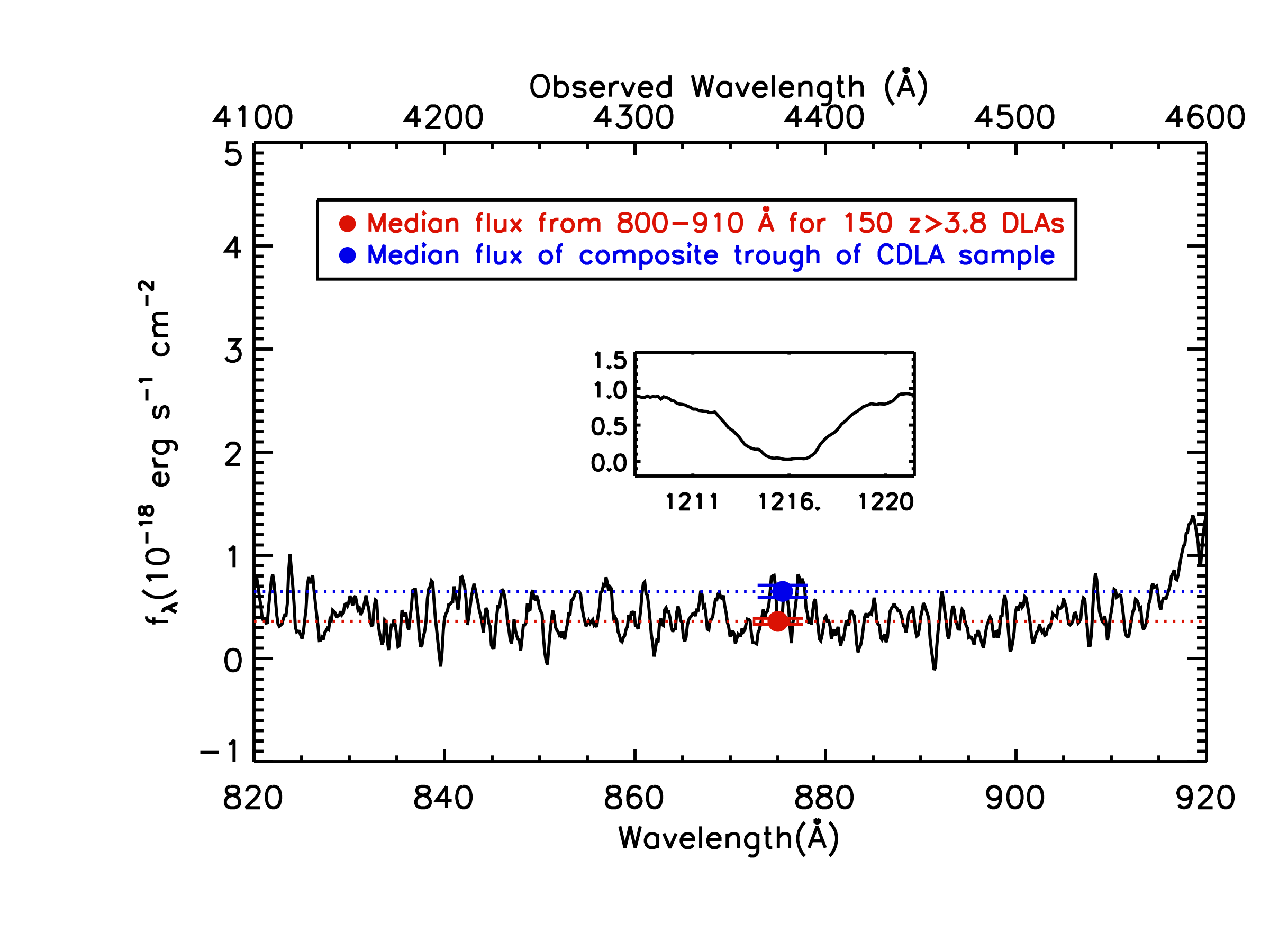}
\caption{Composite spectra of the Lyman limit region (rest-frame 820 \AA - 900 \AA) observed by stacking a group of $150\ z\ge 3.8$ DLAs with a median redshift of $z\sim 4.0$. The inset figure shows a zoom-in of the DLA absorption for the composite spectrum of the group of $z\ge 3.8$ DLAs. The stacked Lyman limit region corresponds to observed frame $\lambda_{\rm{obs}} \sim$ 4100- 4500 \AA. The upper limit of the sky-subtraction can be determined by averaging the flux in the composite Lyman-limit region (see Section 3.2). The average flux of the Lyman limit region (red dot) is $3.5\pm 0.008 \times 10^{-19}$ ergs cm$^{-2}$ \AA$^{-1}$ s$^{-1}$, and the median flux at the composite DLA dark trough (blue dot with errorbar) of the {\it CDLA} sample is $7.6\pm0.6 \times 10^{-19}$ ergs cm$^{-2}$ \AA$^{-1}$ s$^{-1}$, which is significantly larger than the SDSS sky-subtraction residual. The observed wavelength of blue point is determined by DLA center at the median redshift of CDLA sample. }
\end{figure}

 \subsection{Composite spectra in different quasar luminosity bins}
 
In \S3.1, we initially stacked all the spectra of the {\it CDLA} sample, and find the average dark trough flux density is $\bar{F}_{\rm{dark}}= 6.5 \pm 0.6 \times 10^{-19} $ ergs cm$^{-2}$ A$^{-1}$ s$^{-1}$, significantly greater than the upper limit of the sky subtraction residual determined in \S3.2 (Fig.\enskip2). This dark trough flux density indicates that in addition to the sky subtraction residual, flux must be contributed by other sources. In this section, we will examine and test the possible origin of the flux residual in the DLA dark trough. 


 In order to examine the origin of the flux residual in the DLA dark core, we will group the quasars according to their luminosities, and check if the residual intensities in the dark trough correlate with the 
quasar luminosities. The motivation for this experiment is that if the intensity in the stacked dark trough is mainly contributed by the Lyman continua of the foreground DLA, then this intensity should be uncorrelated with the background 
quasar luminosity, and we expect the three luminosity bins to yield similar intensities in the dark trough. 
We divide the quasar into three bins according to their luminosity bins: (1) the highest luminosity bin contains quasars 
with the luminosity density at 1450 \AA\ ($L_{\rm{1450}}) > 9\times10^{42}$ erg s$^{-1}$ \AA$^{-1}$; (2) middle luminosity bin with $5\times10^{42}$  erg s$^{-1}$ \AA$^{-1} < L_{\rm{1450}} < 9\times10^{42}$ erg s$^{-1}$ \AA$^{-1}$; 
(3) low luminosity bin with $L_{\rm{1450}} < 5.0\times10^{42}$ erg s$^{-1}$ \AA$^{-1}$. Each luminosity bin contains 
similar number of quasars. 

The detailed stacking procedure is as follows:  we first calculate the quasar intensity for each quasar using the 
following expression: 
\begin{equation}
 I_{i, \rm{dark}}(\lambda_{\rm{res}})= (f_{\rm{i,sdss}}(\lambda_{\rm{obs}})- f_{\rm{sky}}) 
\times 4\pi {D_L}^2(z_{\rm{QSO}})\times (1+z_{\rm{DLA}}), 
\end{equation}
where $\lambda_{\rm{obs}} =\lambda_{\rm{res}}\times (1+z_{\rm{QSO}})$, $f_{\rm{i,sdss}}$ is the flux density for an individual spectrum, $f_{\rm{sky}}=3.5 \times 10^{-19}$ erg s$^{-1}$ cm$^-2$, 
which is 
determined in \S3.2 and further supported by SDSS calibration group (Schlegel et al. 2014, in prep.). 
Following the stacking method described in \S 3.1, we stack the quasar intensity $I_{i, \rm{dark}} (\lambda_{\rm{res}})$ for each luminosity bin by taking the 3$\sigma$-clipped mean and median value in the dark trough. Outside the dark trough, we just take the median value to stack the spectra.

Note that the $3\sigma$-clipped mean is only a legitimate method when 
we stacking the spectra within the dark trough. In the dark trough, the residual flux, if exists, 
is only in a very low level. It is necessary to first rule out the large outliers due to 
the noise before taking the average. However, outside the dark trough, the different
quasar continua have large scatter in luminosity. The $3\sigma$-clipping will reject large number of bins in the 
continua simply because the quasars differ in luminosity. Therefore, the $3\sigma$-clipped mean 
is no longer a legitimate way to stack the quasar continua outside the dark trough. 
Fig.\enskip3 further demonstrates this point. In Fig.\enskip3, one can see that in the DLA dark trough, the 
number of rejected pixels is small (only $\sim$ 0.5\% of the sample size). However, for the quasar 
continua at rest-frame wavelength $\lambda < 1214$ \AA\ or $\lambda > 1218$\AA, the number of the rejected pixels for 
each wavelength significantly become larger than 1\% of the sample size. This is because outside the 
dark trough, the 3$\sigma$-clipping meanly rejects bins in the continua becasue quasars differ in
luminosity. We have checked that, for the stacked quasar bolometric 
luminosity, the median value is 10\% smaller than the 3$\sigma$-clipped mean, 25\% lower 
than the average without sigma-clipping. Outside the DLA dark trough, we 
take the median to stack the quasar continua. In the next Section (\S 4.2), we will be 
consistent to just use the median composite spectrum to compare the quasar bolometric luminosity
with the dark trough intensities (see details in \S 4.2).

\figurenum{3}
\begin{figure}[tbp]
\epsscale{1}
\label{fig:testschemes}
\plotone{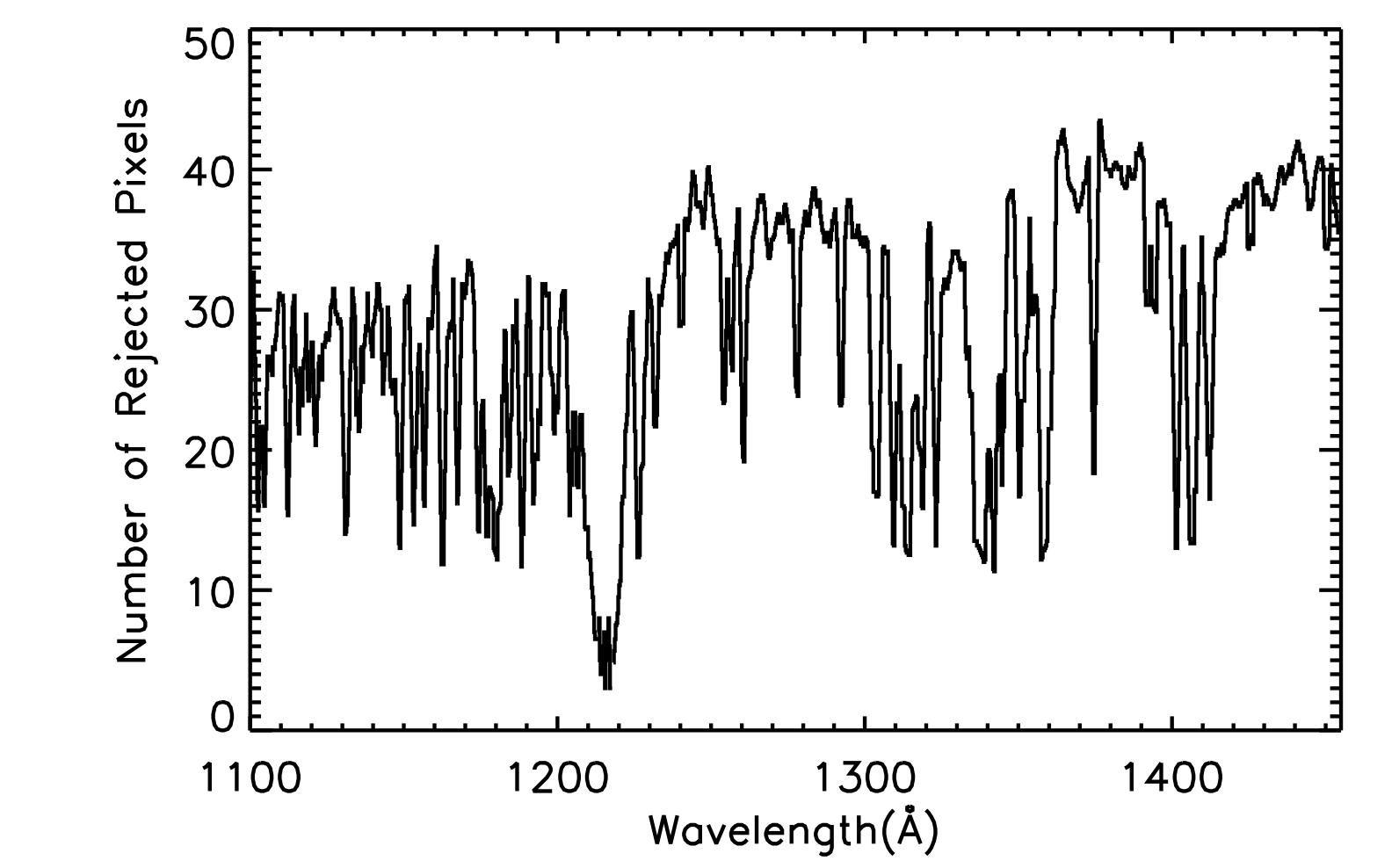}
\caption{We present the number of 3$\sigma$ rejected pixels as a function of rest-frame wavelength for the 
highest quasar luminosity bin. 
From this figure, one can see that in the dark trough region from $\lambda\sim 1214-1218$ \AA, the 
number of 3-$\sigma$ rejected bins is small. On average, 4 pixels at specific wavelength are rejected 
in the dark trough region, about 0.5\% of the sample size (this sumple contains $\sim$ 750 DLAs). This is 
because in the dark trough, the 3$\sigma$-clipping only rejects large outliers due to the noise. However, for the quasar 
continua at $\lambda < 1214$ \AA\ or $\lambda > 1218$ \AA, the number of the rejected pixels is 
significantly larger than 1\% of the sample size. Outside the dark trough, the 3$\sigma$-clipping 
rejects large number of bins in the continua simply because the quasars differ in luminosity. Therefore, outside the 
DLA dark trough, the 3$\sigma$-clipped mean is no longer a legitimate way to stack the quasar continua. 
Outside the DLA dark trough at $\lambda < 1214$ \AA\ and $\lambda > 1218$ \AA, 
we just take the median value to stack the quasar continua (see Section 3.3). }
\end{figure}

 We summarize our results in Table 1 and Fig.\enskip4. The upper panel in Fig.\enskip4 presents the composite DLA spectra in three quasar luminosity bins using the {\it CDLA} sample. The middle panel presents the 3$\sigma$-clipped mean of the DLA dark trough after removing the sky-subtraction residual (Eq. (1)). The stacked dark trough, which lies between the two vertical dashed lines, is defined within $\pm$150 km s$^{-1}$ from the center (see \S 2.2). The lowest panel presents the median value of the flux residual in the DLA dark trough. The zoom-in insets in the middle and lowest panel presents the 3$\sigma$-clipped mean and median value of the overall $CDLA$ sample, centered on the DLA absorption to show the 
non-detection of Ly$\alpha$ emission from the DLA hosts. 
The middle and lower panels demonstrate that the dark trough flux density is non-zero for the two higher luminosity bins, and moreover, the 
dark core intensity increases for the higher quasar luminosity bin. We have compared the results derived from 3$\sigma$-clipped mean and median value. We found the residual intensities derived using 3$\sigma$-clipped mean are consistent with 
the values obtained from the median stacking (Table 1).

\figurenum{4}
\begin{figure*}[tbp]
\epsscale{1}
\label{fig:testschemes}
\plotone{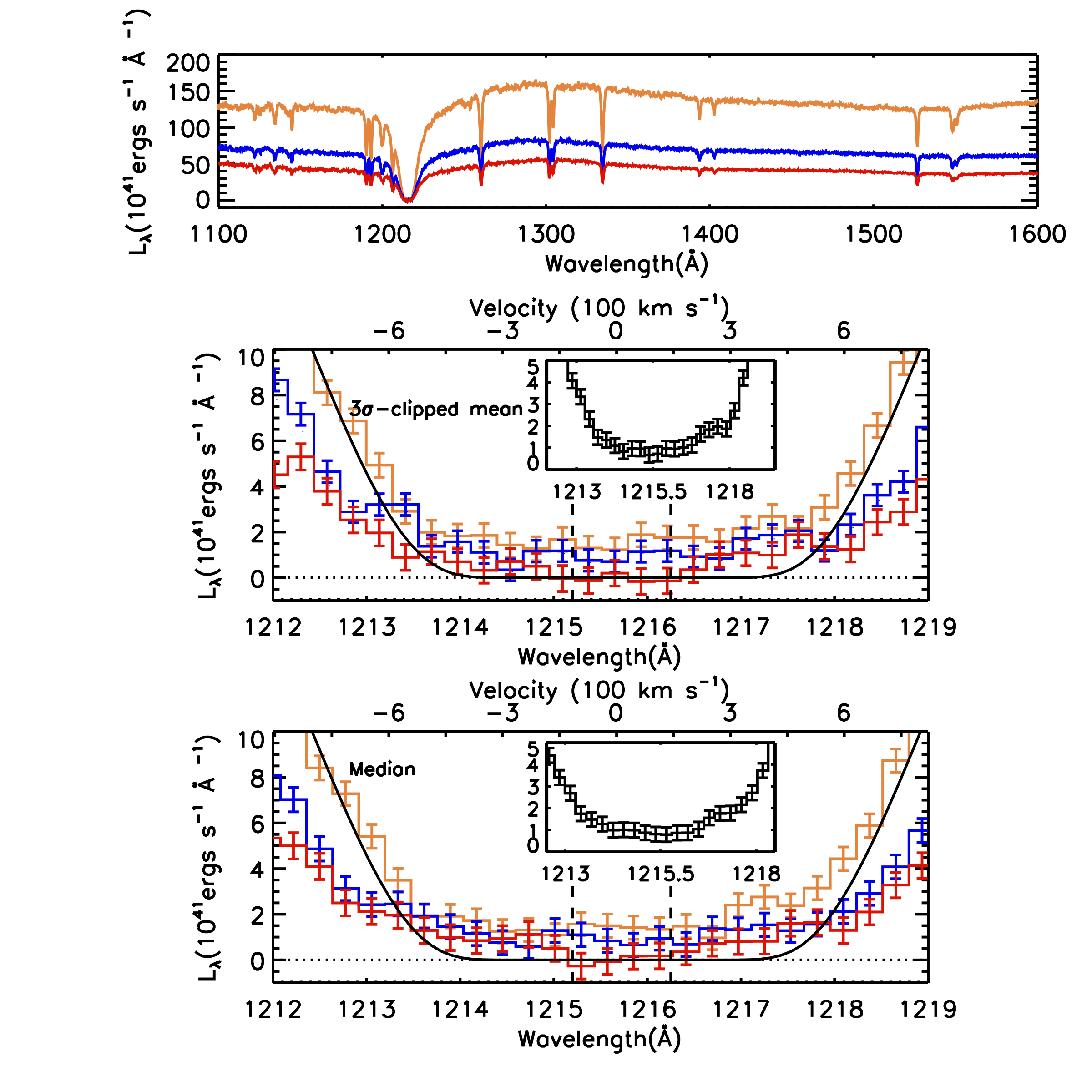}
\caption{The upper panel shows the composite DLA ($N>10^{20.6}$ cm$^{-2}$) spectra for three different luminosity bins of our CDLA sample. The middle panel shows the 3$\sigma$-clipped mean of the expanded DLA dark core region after removing the sky-subtraction residual. The middle and the lower panels demonstrate the statistically significant results that the higher dark core intensity corresponds to the higher quasar luminosity. As stated in (\S2.2), we conservatively define the dark trough within $\pm 150$ km s$^{-1}$ from the center, which region is between the two vertical dashed lines in the middle and lower panel. The insets in the middle and lower panels present the composite spectrum of the full CDLA sample, zoom-in the region around wavelength $\lambda \sim 1216$\AA\ to show the non-detection of the Ly$\alpha$ emission from the DLA host galaxies. The lowest panel shows the median value of the expanded DLA dark core region after subtracting the sky-subtraction residual.}
\end{figure*}

Fig.\enskip5 shows that the correlation between the dark trough intensity and quasar luminosity still holds for DLAs 
with the largest dark troughs in {\it CDLA} sample. DLAs with $N_{\rm{HI}}> 10^{20.9}$ cm$^{-2}$ have an expected dark trough width within 
$\pm 580$ km s$^{-1}$ about the center. 
Again, we conservatively define the dark trough region to be $\pm 250$ km s$^{-1}$ about the center. 
The upper panel of Fig.\enskip5 shows the composite DLA ($N_{\rm{HI}}> 10^{20.9}$ cm$^{-2}$) spectra for the same three luminosity bins with that in Fig.\enskip4. 
The middle panel shows the 3$\sigma$-clipped mean of the expanded DLA dark core region after subtracting the sky-calibration residual, 
and the lowest panel shows the median value for the same dark trough region. The insets in the middle and lower panel present the composite 
spectra of the $N_{\rm{HI}}> 10^{20.9}$ cm$^{-2}$ CDLA sample, centered on the composite DLA absorption to show the non-detection of the 
Ly$\alpha$ emission from the DLA host galaxies. The sample size of $N_{\rm{HI}}> 10^{20.9}$ cm$^{-2}$ {\it CDLA} sub-sample is only half of 
the overall {\it CDLA} sample, yet it is just evident that the higher quasar luminosity bin still corresponds to a higher dark trough intensity, 
i.e., the dark trough intensity with highest quasar luminosity (yellow in Fig.\enskip5) is 3-$\sigma$ higher than that with lowest quasar luminosity 
(red in Fig.\enskip5).

\figurenum{5}
\begin{figure*}[tbp]
\epsscale{1}
\label{fig:testschemes}
\plotone{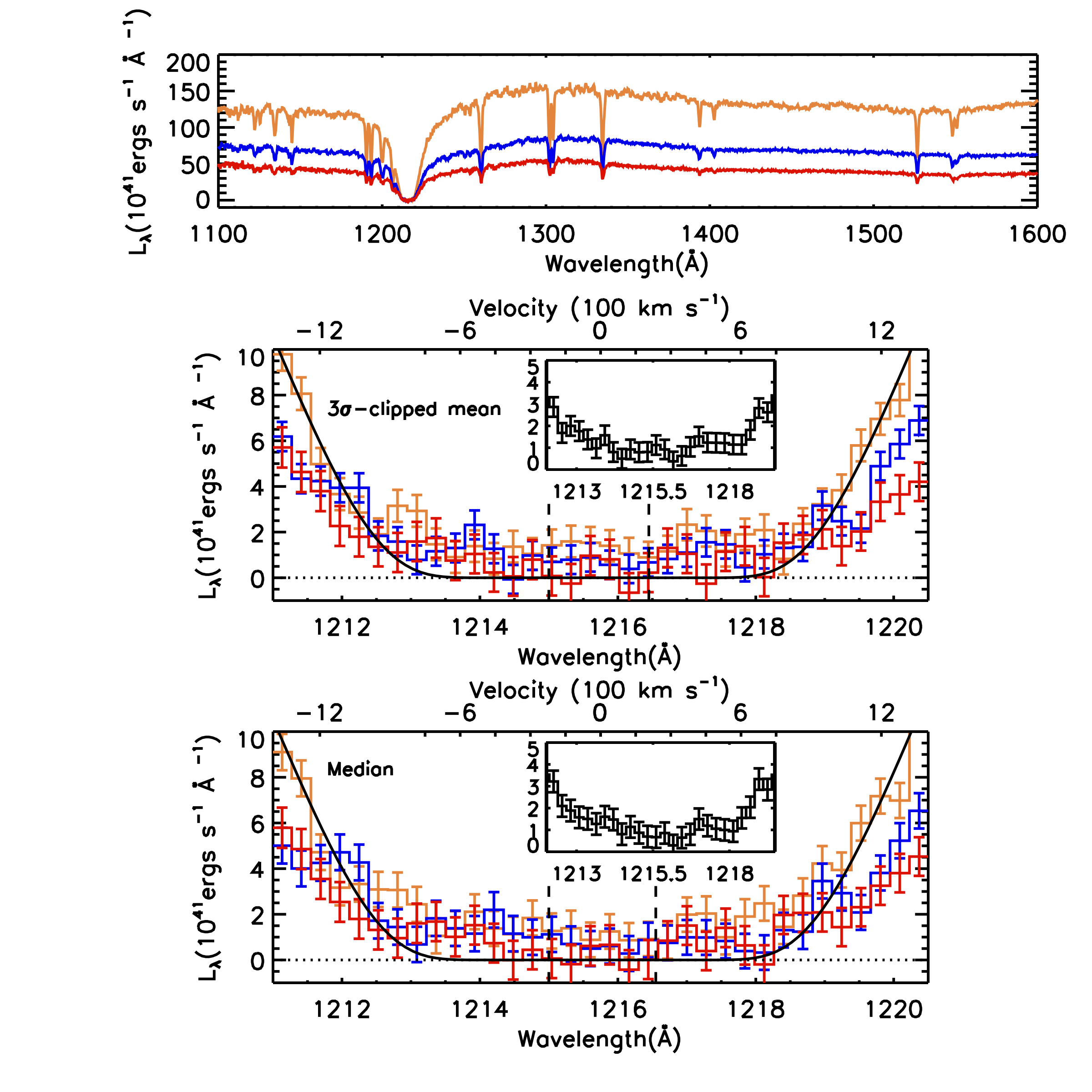}
\caption{The upper panel shows the composite DLA ($N>10^{20.9}$cm$^{-2}$) spectra for three different luminosity bins of the CDLA sample for stronger DLA systems. The middle panel presents the 3$\sigma$-clipped mean of the expanded DLA dark core region after subtracting the sky-calibration residual, and the lowest panel shows the median value of the same dark trough region. In the middle and lower panels, the dark trough is defined between the two vertical dashed lines ($\pm$ 250 km s$^{-1}$ from the center, see \S2.2). The insets in the middle and lower panel present the composite spectrum of $N_{\rm{col}}> 10^{20.9}$ cm$^{-2}$ CDLA sample, centered on the composite DLA absorption to show the non-detection of the Ly$\alpha$ emission from the DLA host galaxies. }
\end{figure*}


The relation between the quasar luminosities and the intensities in the stacked dark trough 
is summarized in Fig.\enskip6. The {\it CDLA} sample with N$_{\rm{HI}}$ $> 10^{20.6}$ cm$^{-2}$ is shown in red, and the
 luminosity distribution is shown in the lower panel in red. A sub-group 
 of DLAs in the {\it CDLA} sample with multiple metal lines are shown in black, where the redshifts are obtained from the low ionization lines. 
The blue points represent DLAs in the {\it CDLA} sample with $N_{\rm{HI}}> 10^{20.9}$ cm$^{-2}$. 
 The comparison between $N_{\rm{HI}}> 10^{20.6}$ cm$^{-2}$ and $N_{\rm{HI}}> 10^{20.9}$ cm$^{-2}$ DLAs demonstrates
 that the dark core intensities are not correlated with the DLA column densities. We also compare DLAs in 
 two bins of equivalent width, which are denoted by triangles in red and yellow color in Fig.\enskip6. We find 
 that the DLA dark core intensities do not correlate with the DLA metal line equivalent width. Fig.\enskip6 clearly 
suggests a $>3\sigma$ detection of the correlation between the mean DLA dark 
 core intensity and quasar luminosity. The histogram in lower panel shows the distribution of the 
 quasar luminosities for $CDLA$ samples.

\figurenum{6}
\begin{figure}[tbp]
\epsscale{1}
\label{fig:testschemes}
\plotone{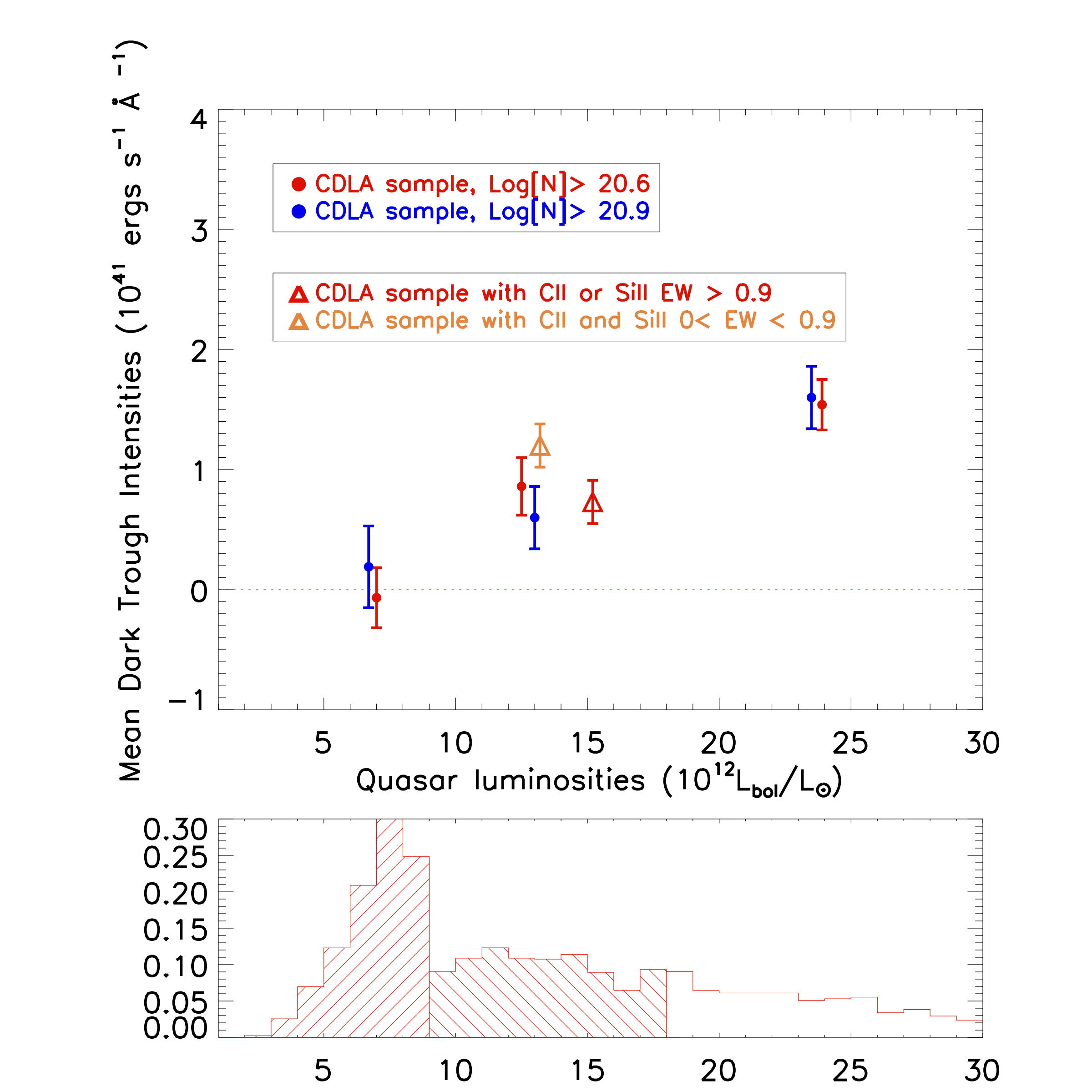}
\caption{A summary of the relation between the quasar luminosities and the intensities in the stacked DLA trough of the CDLA sample. The comparison between DLAs with $N_{\rm{HI}}>10^{20.6}$ (red dots) and $N_{\rm{HI}}>10^{20.9}$ (blue dots) demonstrate that the dark core intensities are not correlated with the DLA column densities. The red and yellow triangles show that the DLA dark core intensities do not correlate with the DLA metal line equivalent width. There is a $\gtrsim 3\sigma$ detection of the correlation between the mean DLA dark core intensities and the quasar luminosities. The histogram in the lower panel shows the distribution of the quasar luminosities for CDLA samples in each of the three luminosity bins.}
\end{figure}

\section{Discussion}

Using the {\it CDLA} sample of 1,940 DLAs, we demonstate that the dark trough intensity is significantly greater than the sky-subtraction residual, and the intensity in the composite dark trough correlates with quasar luminosity. Further, the intensities of the dark trough do not correlate with DLA metal equivalent width or  DLA column density. In this section, we will discuss two possible scenarios of the physical origin for the dark-trough intensity and its relation to the quasar luminosity. 
 
\subsection{DLA galaxy emission}

More than hundreds of DLAs at $z> 2$ have been identified over the last
twenty years. However, only a handful of DLAs have been confirmed to have Ly$\alpha$ emission
originated from the DLA host galaxies. Here, we first give a brief summary of such studies. 
Moller \& Warren (1993) confirmed a Ly$\alpha$ emission 
line associated with a DLA with column density $N_{\rm{col}}$=$10^{21}$ cm$^{-2}$ at $z=2.8$. The 
luminosity of Ly$\alpha$ emission is about $2\times10^{42}$ erg s$^{-1}$ ($\sim 0.4 \times L^*$ at $z=3$) 
(Ciardullo et al. 2012). The velocity offset of the Ly$\alpha$ emission is about +50 $\pm$ 100 km s$^{-1}$. 
Djorgovski et al. (1996) confirmed a Ly$\alpha$ emission associated with
a sub-DLA of column density $N_{\rm{col}}= 10^{20.0}$ cm$^{-2}$ at $z=3.1$. The luminosity of the
Ly$\alpha$ emission is 5$\times$10$^{42}$ erg s$^{-1}$ ($\sim L^*$), and the velocity offset is about +200 km s$^{-1}$. 
Leibundgut \& Robertson (1999) confirmed a Ly$\alpha$ emission from
galaxy associated with a DLA with $N_{\rm{col}}=10^{20.85}$ cm$^{-2}$ at $z=3.1$. The Ly$\alpha$ luminosity is about
$L^*$, and the velocity offset is about 490 km s$^{-1}$. 
Moller et al.(2002) presented a GRB DLA with a $\sim$ 0.5 $L^*$ Ly$\alpha$
emission of DLA galaxy at $z=2.3$, and the
Ly$\alpha$ emission is redshifted by 530 km s$^{-1}$.   
Moller et al. (2004) detect a $\sim$ 0.5 $L^*$ Ly$\alpha$ emission in the dark trough of a DLA 
with large column density $N_{\rm{col}}\sim10^{21.8}$ cm$^{-2}$ at $z=2.0$. The velocity offset of this Ly$\alpha$ emission is about +10 $\pm$ 150 km s$^{-1}$. 
Kulkarni et al. (2012)  presents a super-DLA with column density of
10$^{22.0}$ cm$^{-2}$ associated with a $L^*$ Ly$\alpha$ emission at z=2.2. In the high S/N 
spectrum, the Ly$\alpha$ emission is consistent with a velocity offset of +300 km s$^{-1}$. By stacking a sample of 99 DLAs with $N_{\rm{HI}}> 10^{21.7}$ cm$^{-2}$, Noterdaeme et al. (2014) present the detection of a Ly$\alpha$ emission line with a luminosity of $0.65 \times 10^{42}$ erg s$^{-1}$ in the composite DLA trough, corresponding to 0.1$\times L^*$ galaxies at $z=2-3$, and this composite Ly$\alpha$ emission has a velocity offset of 130 km s$^{-1}$.

After stacking the spectra using the DLAs in the {\it CDLA} sample, we did not detect 
 Ly$\alpha$ emission in the dark trough of the composite spectra (e.g., see the insets in the middle and lower panels of Fig.\enskip4 and Fig.\enskip5).

 In Fig.\enskip7, we present the composite spectra of 2,000 simulated DLAs with Ly$\alpha$ emission being added, and the stacking method is 3$\sigma$-clipped mean. The simulated DLAs have same column density distribution with that of our {\it CDLA} sample. The noise has been included according to the realistic BOSS spectra. The Ly$\alpha$ emission is assumed to have the same luminosity, FWHM (750 km s$^{-1}$) and velocity offset (445 km s$^{-1}$) with the composite spectrum of $L^*$ LBGs at $z=2-3$ (Shapley et al. 2003). Brown line shows the 3$\sigma$-clipped mean of simulated DLAs by assuming that 100\% DLA host galaxies have Ly$\alpha$ emission similar to that of $L^*$ LBGs, and all the Ly$\alpha$ emission entered the SDSS fibers. Blue line assumes 50\% of the simulated DLA host galaxies have $L^*$ Ly$\alpha$ emission. Red line assumes 10\% of the simulated DLA hosts have such strong Ly$\alpha$ emission. Yellow line indicates that the Ly$\alpha$ emission from the DLA hosts do not contribute to the residual flux in the DLA dark trough, either because the emission from DLA host galaxies are below the detection limit (see first probility in \S4.1), or because the impact parameter between the DLA clouds and Ly$\alpha$ emitting region are bigger than 10 kpc (see third probability in \S4.1). From the region between two vertical dashed lines, one can see that the strong Ly$\alpha$ emission will significantly affect the profile of the stacked DLA trough. In the region of the stacked dark trough, with the present of strong Ly$\alpha$ emission, the stacked flux will increases with the wavelength, yet our observed flux (black curve) does not. The observed results are generally consistent with the non-detection of the strong Ly$\alpha$ emission from DLA galaxies. Also, this figure demonstrates that the 3$\sigma$-clipped stacking does not clip away the Ly$\alpha$ emission from the DLA host galaxies.

\figurenum{7}
\begin{figure}[tbp]
\epsscale{1}
\label{fig:testschemes}
\plotone{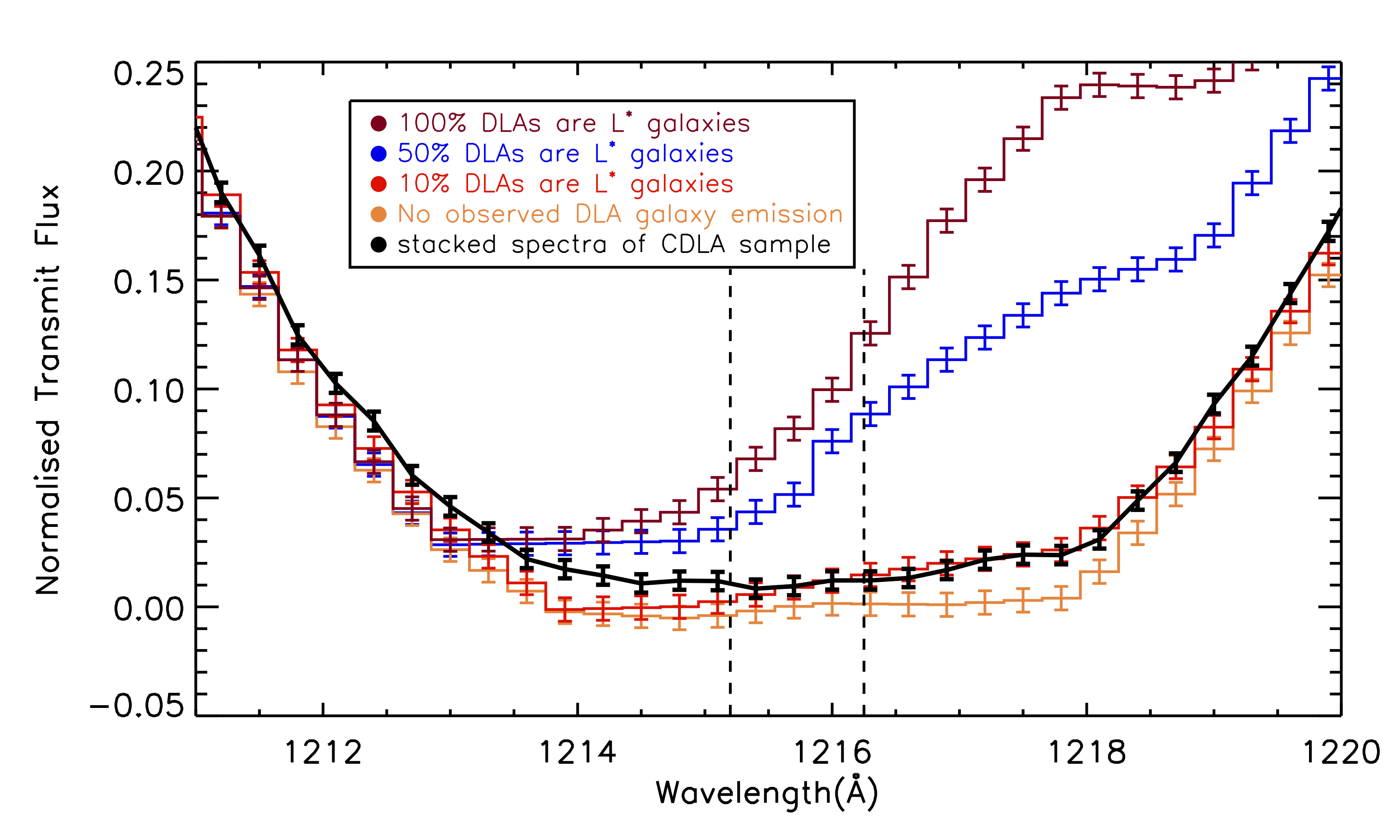}
\caption{The composite spectra of the 2,000 simulated DLAs with realistic galaxy Ly$\alpha$ emission being added. The simulated DLAs are stacked using the 3$\sigma$-clipped mean. Simulated DLAs have same column density distribution with that of our {\it CDLA} sample. The noise has been included according to the realistic observed spectra. The Ly$\alpha$ emission of DLA host is assumed to have the similar luminosity, FWHM and velocity offset with that of $L^*$ LBGs at $z=2-3$ (Shapley et al. 2003).  Brown line shows the stacking of the simulated DLAs by assuming that 100\% of DLA hosts have such strong Ly$\alpha$ emission entering the SDSS fibers. Blue (red) line assumes that 50\% (10\%) of the $L^*$ Ly$\alpha$ emission entering the SDSS fibers for DLA host galaxies. Yellow line indicates none of the emission from the DLA hosts contribute to the residual flux in the DLA dark trough, either because the emission from DLA host galaxies are below the detection limit, or because the impact parameter between the DLA cloud and Ly$\alpha$ emitting region is bigger than 10 kpc (see \S 4.1). The region between two vertical dashed lines indicate the conservative dark trough we defined (same with Fig. 4).  From the region between 1214 \AA\ - 1218 \AA, one can see that the strong Ly$\alpha$ emission will significantly affect the trough profile of the stacked DLAs. With strong Ly$\alpha$ emission, the flux will increase with the wavelength, yet our observed flux (black line) does not. The observed results are generally consistent with the non-detection of the strong Ly$\alpha$ emission from DLA galaxies. Also, this figure demonstrates that the 3$\sigma$-clipped stacking does not clip away the Ly$\alpha$ emission from the DLA host galaxies.}
\end{figure}

Fig.\enskip8 presents a similar plot with Fig.\enskip7. In this figure, we assume that the Ly$\alpha$ emission from DLA host galaxies have the same luminosity with 0.1$\times L^*$ star-forming galaxies at $z = 2-3$ (Noterdaeme et al. 2014). The luminosity, FWHM (300 km s$^{-1}$) and velocity offset (130 km s$^{-1}$) of the Ly$\alpha$ emission are followed by that described in Noterdaeme et al. (2014). Brown color presents the 3$\sigma$-clipped mean of simulated DLAs by assuming that 100\% simulated DLA host galaxies have Ly$\alpha$ emission with the luminosity of 0.1 $L^*$, and all the Ly$\alpha$ emission entered the SDSS fibers. Blue (red) line assumes 50\% (10\%) of the simulated DLA host galaxies have 0.1$\times L^*$ Ly$\alpha$ emission entering the fibers. Yellow line indicates that the flux from the DLA hosts did not contribute to the residual flux in the DLA dark trough (see \S 4.1). From the observed spectra (black line), we did not detect the Ly$\alpha$ emission from DLA host galaxies. Our results are consistent with the above discussions: (1) DLAs are generally hosted by faint galaxies in relatively low-mass halos; (2) for column density of $N_{\rm{col}}> 10^{20.6}$ cm$^{-2}$, most of the DLAs have their impact parameters between the DLA clouds and Ly$\alpha$ emitting regions bigger than 10 kpc.

\figurenum{8}
\begin{figure}[tbp]
\epsscale{1}
\label{fig:testschemes}
\plotone{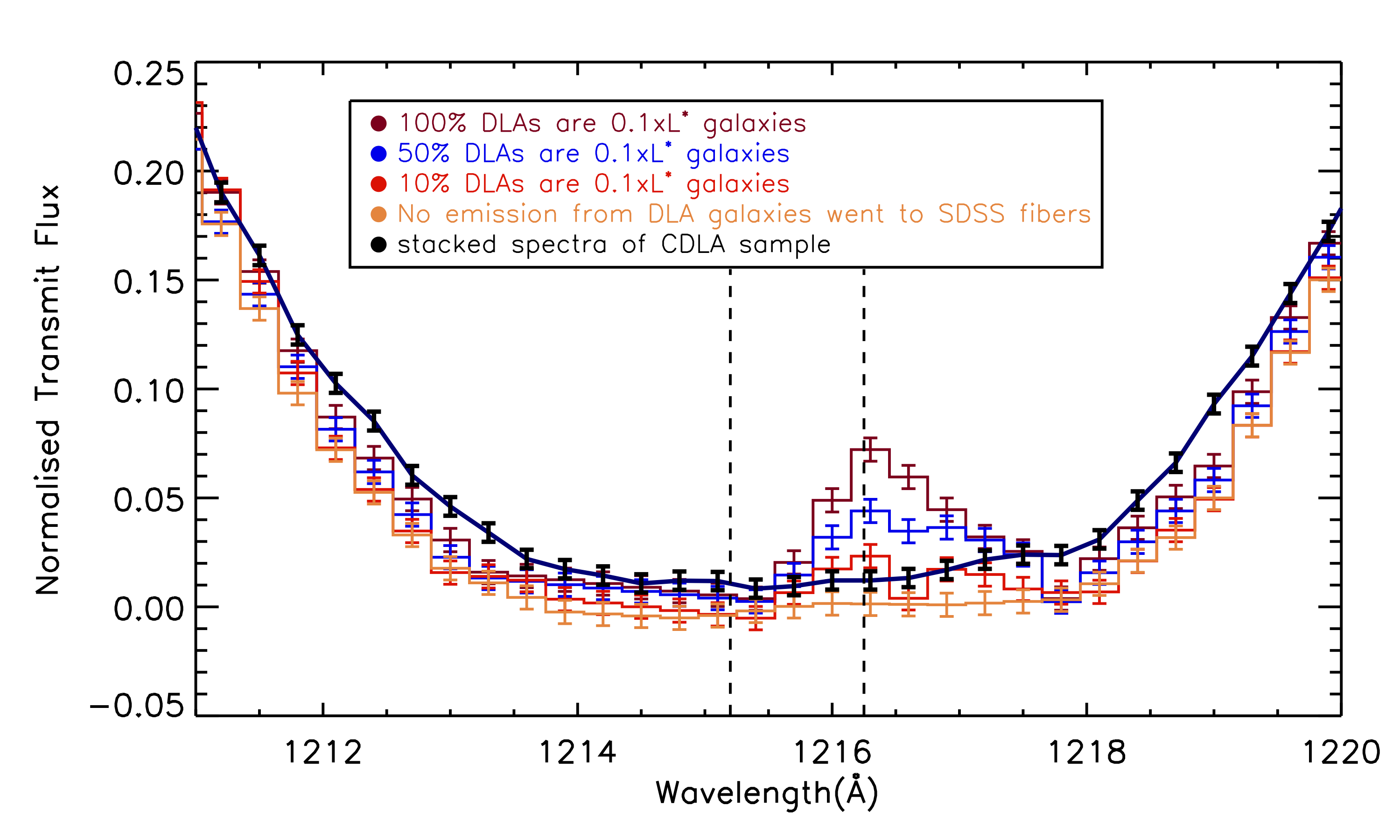}
\caption{Similar plot with Fig.\enskip7. In this figure, the fainter Ly$\alpha$ emission of DLA host galaxies is assumed. We assume the Ly$\alpha$ emission have the similar luminosity, FWHM and velocity offset with that of 0.1$\times L^*$ star-forming galaxies at $z=2-3$ \citep{noterdaeme14}. Brown color presents the 3$\sigma$-clipped mean of simulated DLAs by assuming that 100\% simulated DLA host galaxies have Ly$\alpha$ emission with the luminosity of 0.1$\times L^*$. Blue (red) line assumes 50\% (10\%) of DLAs have 0.1$\times L^*$ Ly$\alpha$ emission.  Yellow line indicates that the flux from the DLA hosts did not contribute to the residual flux in the DLA dark trough (see \S 4.1).  From the observed spectra (black line), we did not detect the Ly$\alpha$ emission from DLA host galaxies. Our results are consistent with the following conclusions (\S 4.1): (1) DLA host galaxies are generally hosted by faint galaxies in relatively low-mass halos; (2) for DLAs with column density of $N_{\rm{col}}>10^{20.6}$ cm$^{-2}$, the impact parameter between the DLA clouds and Ly$\alpha$ emitting regions may be generally bigger than 10 kpc (see \S 4.1). }
\end{figure}

Fig.\enskip9 shows a similar plot with Fig.\enskip8. In this figure, we assume that the Ly$\alpha$ emission from DLA host galaxies have the typical stellar mass of 10$^{8.5}$ M$_\odot$ \citep{moller13}. Assuming the DLA hosts follow the galaxy main sequence, the typical luminosity of DLA host galaxies should be about 0.03$\times L^*$ galaxies at $z=2-3$. In this figure, we further assume that the DLA host galaxies have luminosity of 0.03$\times L^*$. Brown color presents the 3$\sigma$-clipped mean of simulated DLAs by assuming that 100\% simulated DLA host galaxies have Ly$\alpha$ emission with the luminosity of 0.03 $L^*$. Blue (red) line assumes 50\% (10\%) of the simulated DLA host galaxies have 0.03$\times L^*$ Ly$\alpha$ emission. Yellow line indicates that the flux from the DLA hosts did not contribute to the residual flux in the DLA dark trough. From the figure, if only 10\% of the Ly$\alpha$ emission similar to 0.03$\times L^*$, the stacked Ly$\alpha$ emission will below our current detection limit. Black line shows that we did not detect the Ly$\alpha$ emission from DLA host galaxies, which is consistent with the red and yellow lines.

\figurenum{9}
\begin{figure}[tbp]
\epsscale{1}
\label{fig:testschemes}
\plotone{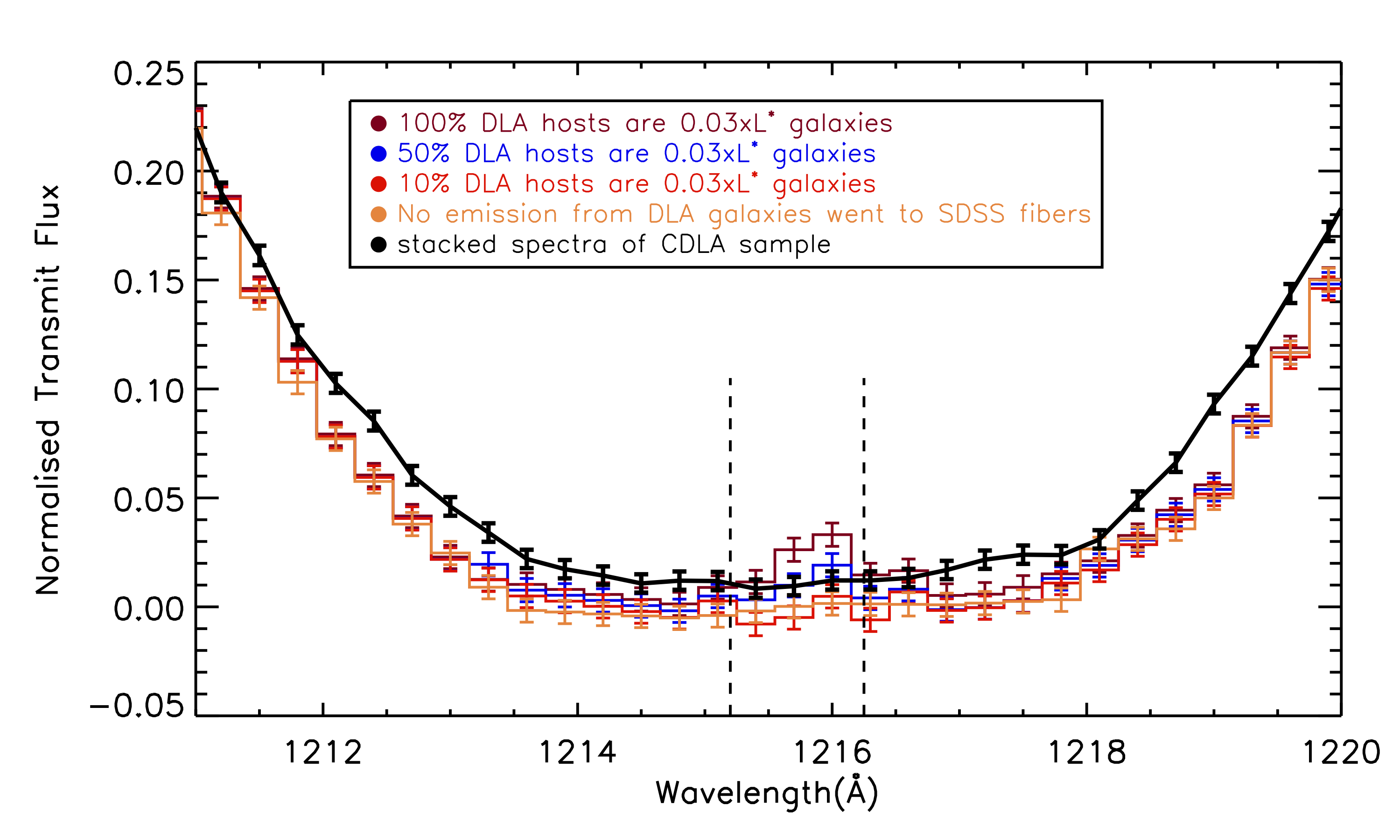}
\caption{In this figure, we assume that the Ly$\alpha$ emission from DLA host galaxies have smaller stellar mass of 10$^{8.5}$ M$_\odot$ \citep{moller13}. Further, if we assume the DLA hosts follow the galaxy main sequence, the typical luminosity of DLA host galaxies should be about 0.03$\times L^*$ galaxies at $z=2-3$. In this figure, we assume that the DLA host galaxies have luminosity of 0.03$\times L^*$. Again, brown color presents the 3$\sigma$-clipped mean of simulated DLAs by assuming that 100\% simulated DLA host galaxies have 0.03$L^*$ Ly$\alpha$ emission entering the SDSS fibers. Blue (red) line assumes that 50\% (10\%) of the simulated DLA host galaxies have Ly$\alpha$ emission with the luminosity of 0.03$L^*$.  Yellow line indicates that the flux from the DLA hosts did not contribute to the residual flux in the DLA dark trough (see \S 4.1). From the figure, if only 10\% of the DLA hosts have Ly$\alpha$ emission with the luminosity of 0.03$L^*$, the stacked Ly$\alpha$ emission is below our current detection limit. From the observed spectra (black), we did not detect the Ly$\alpha$ emission from DLA host galaxies.  }
\end{figure} 

We estimate the 3-$\sigma$ upper limit ($F_{3\sigma}$ (Ly$\alpha$)) by using the gaussian fitting errors, assuming the rest-frame FWHM of the expected Ly$\alpha$ emission is 400 km s$^{-1}$.  The 3-$\sigma$ upper limit of the Ly$\alpha$ emission line flux is $F_{3\sigma}(\rm{Ly}\alpha)= 7.5\times 10^{-19}$ erg s$^{-1}$ cm$^{-2}$. Considering the 2$''$ SDSS-III fiber diameter, this limit translates to a limiting 1-$\sigma$ surface brightness limit of 0.8 $\times 10^{-19}$ ergs s$^{-1}$ cm$^{-2}$ arcsec$^{-2}$. This depth is about a factor of 2 deeper than \citet{rahmani10} and comparable to the depth of the long-slit spectroscopic observations of \citet{rauch08}. At the mean redshift ($\langle z \rangle= 2.6$), the 3-$\sigma$ limit on the Ly$\alpha$ emission corresponds to a Ly$\alpha$ luminosity of 4.0$\times 10^{40}$ ergs s$^{-1}$. 
This upper limit on the Ly$\alpha$ luminosity reaches a depth of $\sim$ 0.01 L$^*$, where 
L$^* \sim 4\times 10^{42}$ erg s$^{-1}$ at $z=3$ \citep{cassata11, ciardullo12}. This corresponds to a 3-$\sigma$ 
upper limit on the Ly$\alpha$-based SFR (SFR$_{\rm{Ly\alpha}}$) of 0.04 M$_\odot$ yr$^{-1}$ 
\citep{kennicutt98, dijkstra10}. 



The non-detection of the Ly$\alpha$ emission from the DLA galaxies in our {\it CDLA} sample can be explained by the following 
possibilities:  (1) DLAs are mostly harboured by faint (sub-$L^*$) galaxies in relatively low-mass dark 
matter halos; (2) A large fraction of 
DLA clouds have impact parameters much larger than the 
stellar half light radius of the DLA host galaxy and exceed the 1$''$ fiber radius, corresponding to $\sim 6$ kpc at $\langle z\rangle =2.65$, such that a large number of the fibers are not exposed to the DLA galaxy stellar light; (3) systemic offsets and redshift 
uncertainties of the Ly$\alpha$ emission lines in the DLA galaxies.

The first possibility is supported by an analysis of the mass-metallicity relation in a sample of 110 DLAs from $z=0.1-5.1$ conducted by \citet{moller13}. They suggest that the typical DLA has a stellar mass of  log($M_*/M_\odot$) $\sim$ 8.5, in agreement with earlier results that most massive DLA galaxies only correspond to the least massive LBGs \citep{fynbo99}. Also, the low-mass DLA halos have been 
suggested by the simulations \cite[e.g.][]{mo99, haehnelt00, nagamine04}. However, a few works favor more massive DLA dark matter halos 
on the order of 10$^{12} $ h$^{-1}$ M$_\odot$ \cite[e.g.][]{font12}. 

  The second possibility is supported by several previous studies. The fibers may not be exposed to the DLA galaxy light due to the large impact parameter of the DLA clouds with respect to the center of the galaxy. \citet{zwaan05} present a comprehensive HI 21cm absorption survey of nearby galaxies, which 
contains 40 HI  absorption features with column densities $N_{\rm{HI}}> 10^{20.0}$ cm$^{-2}$. Half of the DLAs were found to have an impact parameter 
$b> 8$ kpc from the center of the galaxy, indicating that the stellar emission from a large fraction of DLA-hosting galaixes would not enter the 2$''$ BOSS 
fiber, if the DLAs at $z=2-3$ have the same impact parameter as that of the nearby galaxies. \citet{krogager12} studied all of the 10 known 
DLAs with the identified galaxy counterparts at $z=2-3$, and found that DLAs with impact parameter ranges from 0.1$''$- 3.0$''$. 
\citet{pontzen08} are able to reproduce several properties of DLAs in their simulations, and the impact parameters at $z=3$ are expected to 
be $b\lesssim 30$ kpc.
Further, \citet{fynbo08} supports that 
higher metallicity DLAs have larger disks. 
In our {\it CDLA} sample, we only selected DLAs with metal line detections which biases our DLAs to 
higher metallicity, which may have larger HI disks based on the simulations. 
It is true that diffuse halo Ly$\alpha$ emission (Steidel et al. 2011) may enter the fiber. However, the surface brightness of the diffuse Ly$\alpha$ halo emission significantly drops by a factor of $>$3 when the impact parameter is greater than $15$ kpc. 

The third possibility is also supported by numerous observations. Using a sample of 89 Lyman Break Galaxies (LBGs) with 
$\langle z \rangle= 2.3 \pm 0.3$, \citet{steidel10} find that the Ly$\alpha$ lines have systemic velocity 
offsets of $\Delta v_{\rm{Ly\alpha}} = 445 \pm 27$  km s$^{-1}$. Assuming the Ly$\alpha$ emission of DLA galaxy hosts 
has the same systemic velocity offsets as LBGs, the Ly$\alpha$ emission would be located outside of the dark trough of 
DLAs in our {\it CDLA} sample. The three possibilities we have discussed likely work
together to cause the non-detection of DLA Ly$\alpha$ emission. 




By stacking a sample of 99 DLAs with $N_{\rm{HI}}> 10^{21.7}$ cm$^{-2}$, Noterdaeme et al. (2014) present the detection of a Ly$\alpha$ emission line with a luminosity of $0.65 \times 10^{42}$ erg s$^{-1}$ in the composite DLA trough, corresponding to 0.1$\times L^*$ galaxies at $z=2-3$. The idea is that DLAs with large column densities are expected to have smaller galaxy impact parameters. This detection of Ly$\alpha$ emission from DLA galaxies does not contradict our result, because the very large DLAs 
with $N_{\rm{HI}}> 10^{21.7}$ cm$^{-2}$ constitute only $\sim$ 2\% of our {\it CDLA} sample. On the contrary, this further supports the conclusion that most DLAs with $N_{\rm{HI}}< 10^{21.7}$ cm$^{-2}$ are generally not associated with close galaxy counterparts. 
  Our non-detection of Ly$\alpha$ emission supports the hypothesis that the lower column density DLAs could have larger impact parameters (up to a few tens of kpc); thus a large number of damped Ly$\alpha$ absorbers in our sample are located  $>10$ kpc from the central galaxies. 
  
  Besides the contribution of Ly$\alpha$ emission from DLA galaxies, the dark core flux may also be contributed by the DLA Lyman continua at $\lambda \sim 1216$ \AA. However, if the dark core flux comes from the DLA Lyman continua, it is difficult to explain why the dark core flux correlates with quasar luminosity. Therefore, our results favor the interpretation that the flux residual in the dark trough is highly unlikely to be contributed by Lyman continua or Ly$\alpha$ emission from the DLAs.

\subsection{The FUV light from quasar host galaxies}

Theoretically, that the luminous quasar phase naturally coincides with intense star formation was been proposed early on, and has the natural interpretation that both processes rely on reservoirs of gas brought to the center by gas-rich mergers and disk instabilities 
\cite[e.g.][]{sanders88, hopkins06, lutz08}. Several observational studies in the mid-infrared and sub-mm regime \cite[e.g.][]{lutz08, serjeant09} 
confirmed that there exists a link between the quasar host SFR and the black hole accretion rate. 
 \citet{angles13} further propose that galaxy-scale torque-limited accretion \citep{hopkins11} naturally and robustly yields black holes and galaxies evolving along the observed scaling relations, and that AGN feedback does not need to couple with galaxy-scale gas and regulate black hole growth. 
Therefore, FUV stellar light from the quasar hosts is the most plausible interpretation for the flux residual in the composite DLA dark trough. 

It is possible that the DLA absorption, while blocking the quasar continuum region, only partially blocks the host galaxies. The SDSS fiber is exposed to the regions of the quasar host galaxy which are not obscured by the intervening DLA and may contribute residual flux. 
This is consistent with models of clumpy HI distributions in DLA galaxies as suggested by recent observations \cite[e.g.][]{kashikawa13}.
\citet{kanekar09} used the Very Long Baseline Array (VLBA) to image 18 DLAs in redshifted HI 21cm, and suggest that for quasar 
radio emission with source sizes $>100$pc, the median DLA HI covering factor for DLAs at $z=1.5-3.5$ is $f_{\rm{med}}\sim 0.6$. 
If the 21-cm and Ly$\alpha$ (UV) absorption do arise in the same cloud complexes, 
this result indicates that the emission from quasar hosts is not entirely covered by the foreground DLAs. 
Recent simulations and observations also support that in the galactic and circumgalactic environment, the covering factor of optically thick systems is $\lesssim 40 - 50\%$ for $\sim 0.1\ L^*$ galaxies at $z=2-3$ \citep{kanekar09}. 
The half-light radius of quasar host galaxies at $z=2-4$, typically 3-5 kpc (Peng et al. 2006), is fully covered by the SDSS fiber ($\sim 12$ kpc). 
The relation between the dark core residual flux and quasar luminosity supports a correlation between the SFR in the quasar hosts and the quasar luminosity. 

After correcting for the sky-subtraction residual in the highest luminosity bins in the {\it CDLA} sample, the median intensity in the dark core is $\sim 16.4 \pm 2.1 \times 10^{40}$ erg s$^{-1}$ \AA$^{-1}$. We take this value as the average intensity of a quasar host galaxy at a rest-frame wavelength $\lambda \sim 1100$ \AA. This value corresponds to an intensity of $\sim 7.5 \pm 1.0 \times 10^{40}$ erg s$^{-1}$ \AA$^{-1}$ at $\lambda \sim 1700$ \AA\ for a galaxy with a constant star formation history and an age of 0.5 Gyr, after correcting for intergalactic medium absorption \citep{bruzual03, schaye03}. At the mean redshift $\langle z \rangle= 3.1$, this corresponds to 0.3 $L^*$, where $L^*$ is determined from $>$2,000 spectroscopic LBGs from $z= 1.9-3.4$ \citep{reddy08}.  The inferred SFR based on the UV continua (SFR$_{\rm{UV}}$) is $\sim 9$ M$_\odot$ yr$^{-1}$ \citep{madau98}. Following the same procedure, the middle-luminosity bin corresponds to SFR$_{\rm{UV}}$= 5 M$_\odot$ yr$^{-1}$, and the non-detection in lowest luminosity bin  translates to a 3$-\sigma$ upper limit of SFR$_{\rm{UV}}$=7.5 M$_\odot$ yr$^{-1}$. If we assume DLAs have HI covering fraction of $\sim 0.5$ as discussed 
in the last paragraph, the SFRs$_{\rm{UV}}$ for each luminosity bin will be a factor of 2$\times$ higher than our measured 
value. We further derive the bolometric luminosities ($L_{\rm{bol}}$) for the three luminosity bins from the absolute magnitudes at 1450\AA, assuming a bolometric conversion factor $\zeta_{1450\rm{A}}= \frac{L_{\rm{bol}}}{\nu L_{\rm{\nu}, 1450\rm{A}}}= 4.4$ \citep{richards06}. We obtain  $L_{\rm{bol}}= 25.0,\ 13.0,\ 7.0 \times 10^{12}\ L_\odot$, for our three quasar bins; respectively. The properties of these three subsamples are summarized in Table 1 based on our measurements.

\citet{aretxaga98} report that the SFR of quasar host galaxies is $>100-200$ M$_\odot$ yr$^{-1}$ from a sample of three quasars at $z=2$ using $R$ and $I$ band measurements. \citet{jahnke04} estimated a SFR $\sim 2-30\ \rm{M}_\odot$ yr$^{-1}$ for $z=1.8-2.6$ quasar host galaxies using HST ACS F606W and F850LP bands. \citet{villforth08} found a moderate SFRs of $\sim$ 33 M$_\odot$ yr$^{-1}$ for quasar host galaxies using multi-band data from $U$ to $K$. Our SFR$_{\rm{UV}}$ values from partially obscured quasar hosts
 are generally consistent with these previous observations. 

Using the long-slit spectroscopy, Zafar et al. (2011) reported the detection of the host galaxy in the dark trough 
of the DLA in front of quasar Q0151+048A at $z=1.9$. The detected positive residual flux well matches the reported 
brightness of the host galaxy (Fynbo et al. 2000), and the measured counts in the dark trough is about a factor 
of 27 les than the quasar continuum in B-band. In Fig. 10, we compare our results 
with the detection of Zafar et al. (2011). The Ly$\alpha$ mean 
optical depth at $z=2.6$ ($\left< \tau \right>(z=2.6)$) is 0.3;  and $\left< \tau \right>(z=1.9)$ =0.1.  
After correcting the mean Ly$\alpha$ absorption due to IGM, the result in Zafar et al. (2011) is a factor of 2$\times$ 
higher than our results. Note Q0151+48A is one order of magnitude higher than BOSS quasars.  
Also, the detected residual flux for Q0151+48A is the Lyman continuum at $\lambda \sim 1216$ \AA,  
while the residual flux for our BOSS quasar is the Lyman continuum at $\left< \lambda \right> \sim 1100 $ \AA. 
Overall, the difference between our results and that of Zafar et al. (2011) can be well interpreted by the large scatter of the 
covering fraction among different DLA clouds, and plus the uncertainties of the dust extinction 
among different quasar hosts.


Extended scattered nuclear light from the surrounding nebulosity could also be considered as a source for the emission, in addition to light from star formation  \citep{young09}. The signature of scattered nuclear light is suggested by a few polarimetric mesurements \citep[e.g.][]{smith04, letawe07, borguet08}. From Table 1, Fig. 4 and Fig. 5, the residual luminosity we detected is about 1.5\% of quasar luminosity for the highest quasar 
luminosity bin. 
Observations 
of type 2 quasars suggest that scattering efficiencies can be at above or at 1\% level \citep{zakamska05}. With the assumption that any broad Balmer line emission  
that is visible in the off-nuclear spectrum arises from scattered quasar light, \citet{miller03} estimated the scattered light emission from four quasar host galaxies using Keck longslit spectroscopy, and found that the fraction of the scattered nuclear light is generally small compared with the quasar host galaxies' stellar light \citep{miller03}. 
However, \citet{young09} also report that based on their theoretical models, scattered light should be a concern for host characterization in high-redshift observations. From our current observations, any quantitative estimate of the quasar scattered light 
is difficult, and thus,  extended scattered nuclear light remains a possibility for some of the  
residual flux. Thus SFR$_{\rm{UV}}$ we derived could be regarded as a stringent upper limit for the quasar host galaxies.

\figurenum{10}
\begin{figure}[tbp]
\epsscale{1}
\label{fig:testschemes}
\plotone{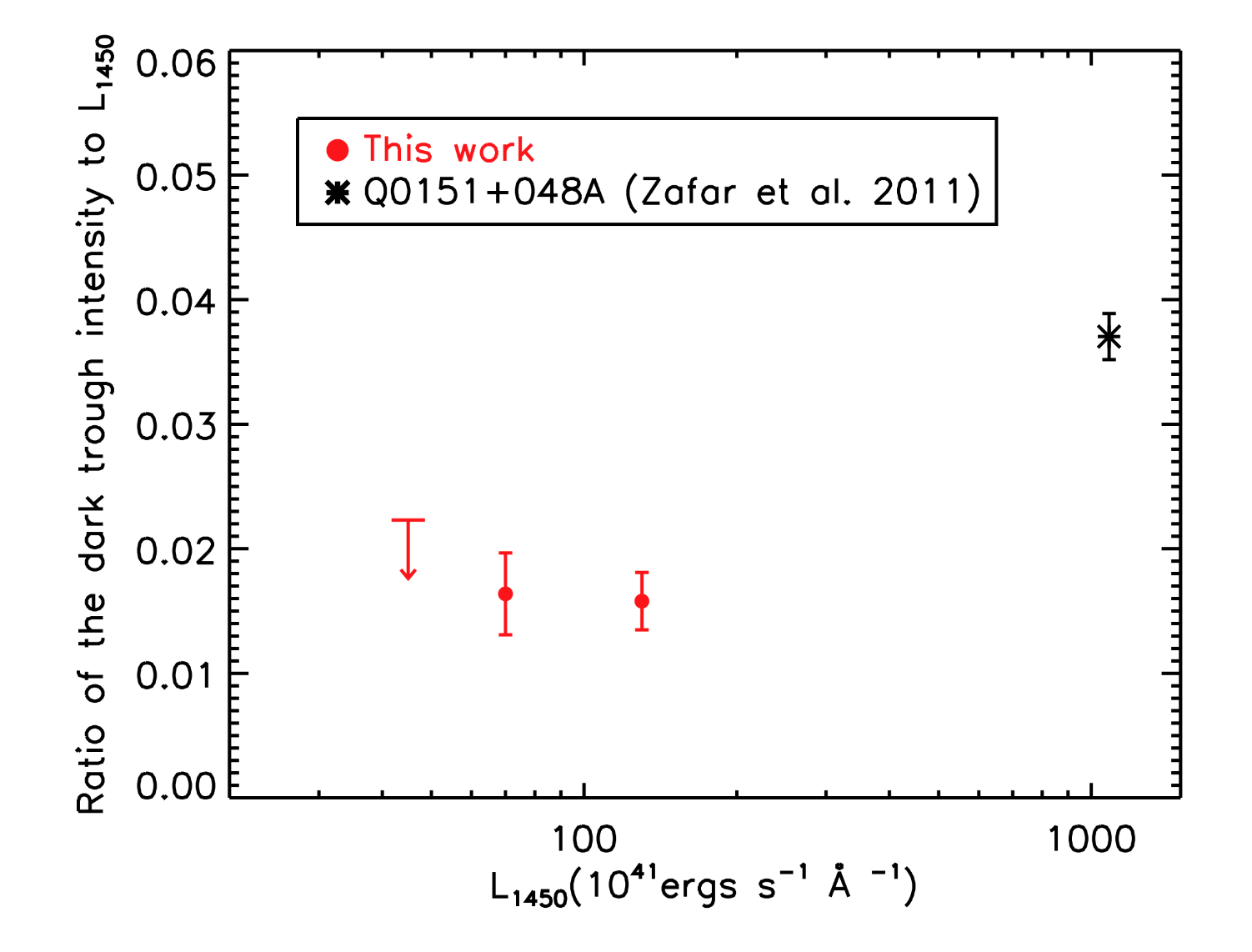}
\caption{ We compare our results (red dots with errorbars)
with the detection of Zafar et al. (2011) (black dot with errorbar). The plot shows the ratio of 
the dark trough residual intensity to the quasar luminosity density at rest-frame 1450 \AA\ ($L_{\rm{1450}}$). 
The dark trough intensity can be regarded as the unobscured FUV emission from the quasar host galaxies (see 
Section 4.2). After correcting the Ly$\alpha$ absorption due to IGM, the result in Zafar et al. (2011) is a factor of 2$\times$ 
higher than our results. Note Q0151+48A is one order of magnitude higher than BOSS quasars.  
Also, the detected residual flux for Q0151+48A is the Lyman continuum at $\lambda \sim 1216$ \AA,  
while the residual flux for our BOSS quasar is the Lyman continuum at $\left< \lambda \right> \sim 1100 $ \AA. 
The difference between our results and that of Zafar et al. (2011) can be well interpreted by the large scatter of the 
covering fraction among different DLA clouds, and plus the uncertainties of the dust extinction 
among different quasar hosts.}
\end{figure}

\section{Summary and Future Prospects}

We define a Clean-DLA ({\it CDLA}) sample from quasars in SDSSIII-BOSS DR10 and stacked all of the spectra in three luminosity bins. In the composite spectra, we do not detect the Ly$\alpha$ emission coming from the DLA host galaxies. We find that the residual intensities in the DLA dark troughs correlate with the quasar luminosities (Fig.\enskip4 and Fig.\enskip5). We further discussed  possible scenarios for the origin of the residual intensity (see details \S4), and conclude that it could be mainly contributed by FUV emission from the quasar host galaxies, rather than sky-subtraction residuals or DLA galaxy emission. For the highest luminosity bin with the median quasar 
bolometric luminosity of 2.5$\times 10^{13} L_\odot$, the median dark trough residual intensity is $= 16.4\pm 2.1 \times 10^{40}$ erg s$^{-1}$ \AA$^{-1}$, correspnding to a 0.3$L^*$ galaxy at $z\sim3$ and an observed SFR$_{\rm{UV}}$ of $\sim$ 9 M$_\odot$ yr$^{-1}$. 
For the middle luminosity bin with the median quasar 
bolometric luminosity of 1.3$\times 10^{13} L_\odot$, the median residual intensity in the stacked 
dark trough is $= 8.8\pm 2.3 \times 10^{40}$ erg s$^{-1}$ \AA$^{-1}$, correspnding to a 0.2$L^*$ galaxy at $z\sim3$ and an observed SFR$_{\rm{UV}}$ of $\sim$ 5 M$_\odot$ yr$^{-1}$. 
We do not detect the residual intensity in the stacked dark trough for the lowest bolometric luminosity. 
The median quasar bolometric luminosity for the lowest luminosity bin is 7.0$\times 10^{12} L_\odot$, 
and we put a 3-$\sigma$ upper limit on dark trough intensity of 7.5$\times 10^{40}$ erg s$^{-1}$ \AA$^{-1}$, and a 3-$\sigma$ upper limit on SFR of 4.5 M$_\odot$ yr$^{-1}$.


The more comprehensive study of the quasar host galaxies should combine the data from the rest-frame UV to rest-frame FIR. The SFR$_{\rm{UV}}$ we derived may not represent a complete measure of the energy generated from high-redshift quasar hosts due to dust obscuration. Models suggest an evolutionary link between the optical quasar and local ULIRGs (SMGs at high redshift) (e.g. Sanders et al. 1988), where quasars are proposed to emerge from SMGs. Thus, it is possible that quasar host galaxies may still be heavily enshrouded by dust, and that a considerable amount of the UV emission from star formation is dust-obscured.  FIR emission should 
provide a more comprehensive measure of the total SFR in dusty circumnuclear starbursts, and a direct comparison 
between SFR$_{\rm{UV}}$ and SFR$_{\rm{FIR}}$ provides an estimate for the dust obscuration in the quasar host galaxies. 

However, quasars at $z=2-5$ with the similar bolometric luminosities with BOSS quasars
have not been observed in the rest-frame FIR. With ALMA, one can probe a carefully-selected 
sample of BOSS quasars at $z\sim 2-3$ with reasonable exposure times. The 
previous studies have determined a relation between the bolometric luminosities ($L_{\rm{bol}}$) and  rest-frame FIR luminosity ($L_{\rm{FIR}}$), using quasars at higher redshifts ($z\gtrsim 5$)
and quasars at $z=2-4$ but with much higher bolometric luminosities than BOSS quasars \citep{wang11, omont03, omont01, carilli01, priddey03}. Assuming this relation between $L_{\rm{bol}}$ and $L_{\rm{FIR}}$ can be applied to quasars with similar luminosity to our sample at $z=2-3$, we can predict that for our highest luminosity bins, the $L_{\rm{bol}}$-inferred $L_{\rm{FIR}}$ is $\sim 10^{12.4}\ L_{\odot}$. 
Using FIR/sub-mm spectral energy distribtuions (SEDs), previous studies suggest that a significant fraction of the rest-frame
FIR emission comes from massive star formation, possibly indicating
the formation of early galactic bulges (Leipski et al. 2012;
2014). Several studies also simply suggest 50\% of the FIR emission is powered by the star
formation (e.g., Bertold et al. 2003; Wang et al. 2008, 2011). If we also assume 50\% of the $L_{\rm{bol}}$-inferred $L_{\rm{FIR}} \sim 10^{12.4}\ L_{\odot}$ is powered by star formation, 
the median FIR-derived star formation rate (SFR$_{\rm{FIR}}$) for our highest quasar luminosity bin should be $\sim$ 300 M$_\odot$ yr$^{-1}$. This expected SFR$_{\rm{FIR}}$ is already more than one orders of magnitude higher than the SFR$_{\rm{UV}}$ we derived. 

Future detailed 
studies from rest-frame IR to sub-mm will constrain better dust heating models of 
 host galaxies for BOSS quasars. The next generation spaced-based and ground-based telescope adaptive optics will provide significantly higher spatial resolution and enable 
much more precise PSF subtraction, and make it possible to fully resolve and accurately measure the UV emission from quasar hosts at $z\gtrsim 2$. 
A direct comparison between the observed SFR$_{\rm{FIR}}$ and 
SFR$_{\rm{UV}}$ will directly probe the dust obscuration in the quasar host galaxies, and can be 
used to compare with the quasar evolutionary models.


{{\it  Acknowledgement:} We thank the anonymous referee for insightful comments which have significantly improved the paper. ZC thanks George Becker and J. Xavier Prochaska for useful discussions. ZC, XF acknowledge support from NSF grants AST 08-06861 and AST 11-07682. Funding for SDSS-III has been provided by the Alfred P. Sloan Foundation, the Participating Institutions, the National Science Foundation, and the U.S. Department of Energy Office of Science. The SDSS-III web site is http: //www.sdss3.org/. SDSS-III is managed by the Astrophysical Research Consortium for the Participating Institutions of the SDSS-III Collaboration including the University of Arizona, the Brazilian Participation Group, Brookhaven National Laboratory, University of Cambridge, Carnegie Mellon University, University of Florida, the French Participation Group, the German Participation Group, Harvard University, the Instituto de Astrofísica de Canarias, the Michigan State/Notre Dame/JINA Participation Group, Johns Hopkins University, Lawrence Berkeley National Laboratory, Max Planck Institute for Astrophysics, Max Planck Institute for Extraterrestrial Physics, New Mexico State University, New York University, Ohio State University, Pennsylvania State University, University of Portsmouth, Princeton University, the Spanish Participation Group, University of Tokyo, University of Utah, Vanderbilt University, University of Virginia, University of Washington, and Yale University. }


   
%
%

%
%
\paragraph{}


\clearpage

\begin{deluxetable*}{lcccc}  
\tablecolumns{6}
\tablewidth{0pt}
\tablecaption{Dark Trough Intensities for different Quasar Luminosity Bins}
\label{table:F130N_S}	
\tablehead{\colhead{$I_{\rm{QSO}}$ ($\lambda= 1450 \rm{\AA}$)} & 
	          \colhead{$L_{\rm{bol}}$ } & 
                  \colhead{$I_{\rm{trough}}$ (median) } &
                  \colhead{$I_{\rm{trough}}$ ($3\sigma$-clipped mean) } &
                  \colhead{SFR$_{\rm{UV}}$ }  \nl
(erg s$^{-1}$ \AA$^{-1}$) & ($10^{12} L_\odot$) & (erg s$^{-1}$ \AA$^{-1}$) & (M$_\odot$ yr$^{-1}$) & (M$_{\odot}$ yr$^{-1}$) }
\startdata
 $1.3 \times 10^{43}$ & 25.0 & $16.4 \pm 2.1 \times 10^{40}$ & $15.2\pm 2.1 \times 10^{40} $ & 9  \nl
 $6.6 \times 10^{42} $ &  13.0 & $8.8 \pm 2.3 \times 10^{40}$ & $8.2 \pm 2.3\times 10^{40}$    & 5  \nl
 $4.4\times 10^{42}$  &  7.0 & 3-$\sigma < 7.5 \times 10^{40}$  &  3-$\sigma<7.5\times 10^{40}$ &  3-$\sigma < 4.5$ \nl
\enddata
\footnotetext[1]{The intensities $I_{\rm{trough}}$ are calculated using Eq. (1) which mean the sky-subtraction residual has been corrected}
\end{deluxetable*}
\end{document}